\documentclass[12pt,headings=big,numbers=noenddot,DIV=14,a4paper]{article}%

\pdfoutput=1

\usepackage{amsmath,amssymb,amsfonts}
\usepackage[normal,font=small,labelfont=bf,labelsep=period]{caption}
\usepackage[pdftex]{color,graphicx}
\usepackage[english]{babel}
\usepackage[compress]{cite}

\usepackage[dvipsnames]{xcolor}
\definecolor{darkgreen}{rgb}{0,0.65,0}
\usepackage{slashed}

\usepackage[linktoc=page,bookmarks=false,colorlinks=true,
linkcolor=WildStrawberry,citecolor=BlueGreen,urlcolor=RedViolet]{hyperref}

\numberwithin{equation}{section}
\DeclareFontFamily{OT1}{pzc}{}
\DeclareFontShape{OT1}{pzc}{m}{it}{<-> s * [1.10] pzcmi7t}{}
\DeclareMathAlphabet{\mathpzc}{OT1}{pzc}{m}{it}

\newcommand{\met}{\ensuremath{{\not\mathrel{E}}_T }}

\begin{document}

\begin{flushright}
\footnotesize
SACLAY--T14/102
\end{flushright}
\color{black}

\begin{center}

{\huge \bf Wino-like\\[0.2cm] Minimal Dark Matter\\[0.6cm]
      and future colliders}

\medskip
\bigskip\color{black}\vspace{0.5cm}

{
{\large\bf Marco Cirelli}\ $^a$,
{\large\bf Filippo Sala}\ $^{a}$,
{\large\bf Marco Taoso}\ $^{a}$
}
\\[7mm]
{\it $^a$ \href{http://ipht.cea.fr/en/index.php}{Institut de Physique Th\'eorique}, CNRS, URA 2306 \& CEA/Saclay,\\ 
	F-91191 Gif-sur-Yvette, France}\\[3mm]
\end{center}

\bigskip

\centerline{\bf Abstract}
\begin{quote}
\color{black}
We extend the Standard Model with an EW fermion triplet, stable thanks to one of the accidental symmetries already present in the theory. 
On top of being a potential Dark Matter candidate, additional motivations for this new state are the stability of the vacuum, the fact it does not introduce a large fine-tuning in the Higgs mass, and that it helps with gauge coupling unification.
We perform an analysis of the reach for such a particle at the high-luminosity {\sc Lhc}, and at a futuristic 100 TeV $p p$ collider. 
We do so for the monojet, monophoton, vector boson fusion and disappearing tracks channels. At 100 TeV, disappearing tracks will likely probe the mass region of 3 TeV, relevant for thermally produced Dark Matter. The reach of the other channels is found to extend up to $\sim$ 1.3 (1.7) TeV for 3 (30) ab$^{-1}$ of integrated luminosity, provided systematics are well under control.
This model also constitutes a benchmark of a typical WIMP Dark Matter candidate, and its phenomenology reproduces that of various models of Supersymmetry featuring a pure Wino as the lightest sparticle.
\end{quote}




\newpage

\tableofcontents

\section{Introduction}
\label{Introduction}

The first run of the {\sc Lhc} ended without any direct evidence for New Physics (NP). 
Analogously, precision measurements of flavour, electroweak and Higgs observables have all shown a remarkable agreement with the Standard Model (SM) predictions.
This situation casts doubts on naturalness of the Fermi scale, at least in its stricter sense~\cite{'tHooft:1979bh}, as \textit{the} guiding principle to build NP models and, ultimately, to provide a guidance for future experimental searches.

Other solutions to the hierarchy problem exist, that are not weakened by the absence of physics beyond the Standard Model (BSM) at current experiments, nor eventually at future ones. 
One amounts to loose the requirement of naturalness, by assuming that gravity does not give large radiative corrections to the Higgs mass~\cite{Yoon:2002nt,Shaposhnikov:2007nj}.
While the viability of such an assumption is still a matter of inquiry~\footnote{See e.g.~\cite{Dubovsky:2013ira} for a two dimensional example and~\cite{Giudice:2013yca} for a more detailed discussion of this possibility.}, it opens interesting new avenues in model building: to be allowed, any NP has at least not to give large contributions to the Higgs mass~\cite{Farina:2013mla}.
Another solution consists in letting the parameters of the theory scan, in some sense, over an ensemble (a ``Multiverse'') of values, on which the measured ones are selected by a criterion, like the anthropic one (see \cite{Wilczek:2013lra} for a recent discussion).
Both these possibilities, at present, do not give by themselves any indication of where to expect NP to show up.

Such an indication may be provided, at the price of losing generality, by specific solutions to other problems of the Standard Model, like for example the nature of Dark Matter (DM). In this paper we will explore such a case, motivating an {\em electroweak fermion triplet} as a minimal candidate for weakly interacting massive particle (WIMP) Dark Matter. A mass in the multi-TeV range naturally arises in this scenario, which poses a challenge to detection. We will review the status of Direct and Indirect Detection searches, and explore in detail the phenomenology at future colliders.

With this last respect, the analysis we present constitutes a useful benchmark case. In fact, till now {\sc Atlas} and {\sc Cms} have mainly cast their searches for Dark Matter using an effective field theory language, also to allow for a simple comparison with the limits from Direct Detection experiments. Given the high center-of-mass energies, such a choice is at least questionable for the 8 TeV run of the {\sc Lhc}, and its domain of validity will further shrink at higher energies~\cite{Busoni:2013lha,Busoni:2014sya,Busoni:2014haa,Buchmueller:2013dya}. New ways of running the searches and
expressing the bounds will be a necessity, the main options being simplified and, indeed, benchmark models (see e.g.~\cite{Shoemaker:2011vi,Dreiner:2013vla,Chang:2013oia,An:2013xka,Bai:2013iqa,DiFranzo:2013vra,Papucci:2014iwa,deSimone:2014pda} for previous examples).

In particular, we will compute the exclusion reaches of the {\sc Lhc-14} with an integrated luminosity L~=~3 ab$^{-1}$, and of a 100 TeV $pp$ collider, for L~=~3 and  30 ab$^{-1}$. The project of such a high energy machine is under thorough discussion in the community, and we believe this study adds to the physics motivations for its realisation.

\medskip

The rest of the paper is organized as follows. 
In Section~\ref{sec:motivations} we introduce explicitly the model and illustrate the reasons why it is interesting, as a DM candidate and beyond.
Section~\ref{sec:collider} contains our main results for the monojet, monophoton, vector boson fusion and disappearing tracks channels. We present in detail the analyses we have performed, as well as the link with previous literature on the subject.
In Section~\ref{sec:DDIDpheno} we briefly address the phenomenology of such a candidate concerning Direct and Indirect DM searches. 
In Section~\ref{sec:conclusions} we summarize and conclude.


\section{The model: construction and motivations}
\label{sec:motivations}
We add to the Standard Model particle content a fermion $\chi$, triplet under the $SU(2)_L$ group and singlet under color and hypercharge ($Y=0$). We insist that all the possible interactions of $\chi$ with Standard Model particles have to preserve the gauge and accidental symmetries of the Standard Model, i.e. most notably lepton number or $B-L$\footnote{or even a discrete subgroup, like matter parity.}, under which $\chi$ is assumed to be neutral. This last requirement is crucial since it 
forbids the presence of higher dimensional operator that could lead to a decay of $\chi$.
Hence, the phenomenologically relevant Lagrangian is very simple and reads

\begin{align}
\label{eq:Lagrangian}
\mathcal{L}_\chi &= \frac{1}{2}\,\bar{\chi}(i \slashed{D} - M_\chi) \chi \nonumber \\
& = \frac{1}{2}\,\bar{\chi_0}(i \slashed{\partial} - M_{\chi_0}) \chi_0 +\bar{\chi^+}(i \slashed{\partial} - M_{\chi^\pm}) \chi^+ \\
&+ g \left( \bar{\chi^+}\gamma_\mu \chi^+ (s_w A_\mu +c_w Z_\mu) + \bar{\chi^+}\gamma_\mu \chi_0 W_{\mu}^- + \bar{\chi_0}\gamma_\mu \chi^+ W_{\mu}^+
\right) \nonumber 
\end{align}
where $g$ is the $SU(2)$ gauge coupling, and $s_w$ and $c_w$ are the sine and cosine of the Weinberg angle. The difference $M_{\chi^\pm}-M_{\chi_0}$ at the two-loop level is $164\div165$ MeV (stable to the level of 1 MeV for $M_{\chi_0} \gtrsim 1 $ TeV)~\cite{Ibe:2012sx}.

\bigskip

This minimalistic picture is directly inspired by the Minimal Dark Matter model~\cite{Cirelli:2005uq,Cirelli:2009uv}, which had already considered the phenomenology of EW multiplet as DM candidates. In that construction, however, the main focus had been dedicated to the 5-plet particle, which does not require the enforcement of $B-L$ for stability. 
The triplet under examination here, on the other hand, possesses several virtues that make it very interesting, even beyond the DM motivation.  
Let us schematize the main different reasons, both phenomenological and theoretical:
\begin{itemize}

\item[$\diamond$] With the enforcement of $B-L$, $\chi$ is automatically stable, making it a potential candidate to constitute part or all of the Dark Matter. \\
More precisely, if one requires that $\chi$ is thermally produced, via the standard freeze-out mechanism, and that it constitutes the whole of DM, then its mass $M$ is univocally determined to be $M \simeq 3.0 \div 3.2$ TeV (we adopt here the value from~\cite{Hryczuk:2014hpa}, which takes into account all higher order corrections, including the Sommerfeld enhancement).\\
Other ranges of masses, however, are also interesting. For $M \lesssim 3$ TeV, $\chi$ is a subdominant DM component if thermally produced, or it can still be the whole of DM if a non-thermal production history is assumed. For this region of mass, as we will see, collider searches are possible. For $M \gtrsim 3.2$ TeV non-thermal production has to be assumed to avoid the over closure of the universe. \\
In the following we will leave $M$ as a free parameter. 

\item[$\diamond$] The presence of an EW multiplet changes the running of the Higgs quartic coupling ~\cite{Chao:2012mx}, increasing its value at higher energies. This helps to push the Higgs potential towards the stability regime, making it less uncomfortable. Moreover, the recent {\sc Bicep2} discovery~\cite{Ade:2014xna}, if confirmed, suggests that the EW vacuum in which we live in would have already decayed~\cite{Espinosa:2007qp,Kobakhidze:2013tn,Fairbairn:2014zia,Enqvist:2014bua,Kobakhidze:2014xda,Hook:2014uia} in a universe where the SM holds up to the instability scale $\Lambda = 10^{9\div11}$ GeV.
In this case, unless Planck suppressed corrections stabilise the vacuum during inflation~\cite{Hook:2014uia}, some new physics making the quartic Higgs coupling larger than zero would be needed. The introduction of an EW fermion triplet is one of the simplest possibilities (see e.g.~\cite{EliasMiro:2012ay,Chao:2012mx}) that address this issue.

\item[$\diamond$] The same EW triplet changes also the running of the gauge couplings, making $g_1$ and $g_2$ unify, at one loop, at a scale of $\simeq 10^{15}$ GeV (see e.g.~\cite{Giudice:2004tc, Frigerio:2009wf}). It is remarkable that the triplet is the only $SU(2)_L$ fermion multiplet that allows, if alone, such scale to lie between M$_{\rm GUT}$ and M$_{\rm Planck}$~\cite{Frigerio:2009wf}. Concerning two loops, the multiplet which suffers from the most severe one-loop cancellations is the quintuplet, and the two-loop corrections are expected to worsen its situation~\footnote{M.~Nardecchia et al., private communication, paper to appear.}. For the triplet they are expected to raise by a factor of $\sim$ 2 the scale of unification of $g_1$ and $g_2$~\cite{Frigerio:2009wf}. In any case, such a value for the scale of a grand unified theory, in its simplest realisations, would be already excluded by the severe bounds on proton decay (again, see e.g.~\cite{Giudice:2004tc}). However it is not difficult to imagine theoretical constructions that avoid this problem (see e.g.~\cite{Frigerio:2009wf} or~\cite{Mahbubani:2005pt}).

\item[$\diamond$] This minimal model has some interest also in relation to the hierarchy problem of the Fermi scale, 
if one assumes that gravity does not influence its radiative stability, like proposed in~\cite{Shaposhnikov:2007nj,Yoon:2002nt} and recently reelaborated upon in~\cite{Farina:2013mla,Dubovsky:2013ira}. In fact values of $M_\chi \leq 3$  TeV would imply a fine-tuning in the Higgs mass value at a level of 10\% or better~\cite{Farina:2013mla}, more precisely $M_{\chi} < 1.0$ TeV$\times \sqrt{\Delta}$, with $\Delta$ the amount of fine-tuning. Notice that, to achieve the same small amount of fine-tuning, larger EW multiplets (like a quintuplet) would have to be much lighter.
This adds to the motivation for a `light' $\chi$.

\item[$\diamond$] Last but not least, this minimal model can be seen as a benchmark of the typical thermal-relic WIMP Dark Matter candidate, as we already mentioned. It is a prototype of more complicated models, and it can reproduce their low energy phenomenology to a remarkable accuracy.  For example it effectively reproduces more complete unified models like~\cite{Frigerio:2009wf}, as well as a Supersymmetric spectrum in which a pure Wino is the lightest EW superpartner, a possibility that recently attracted lot of attention in different models of SUSY at higher scales~\cite{Wells:2003tf,ArkaniHamed:2004fb,Giudice:2004tc,Arvanitaki:2012ps,Hall:2011jd,Hall:2012zp,Hall:2013eko,Hall:2014vga}. 
Here it is important to stress that the value of the mass splitting $M_{\chi^\pm}-M_{\chi_0}$, which is crucial for the phenomenology we will discuss, is robust against corrections from possible UV models. In fact, as remarked in~\cite{Ibe:2012sx}, the first operator that can induce a further splitting arises at dimension~7.
\end{itemize}


\section{Detection at 14 and 100 TeV $pp$ colliders}
\label{sec:collider}

In this Section we analyze the prospects for detection of the Wino-like Dark Matter candidate at proton-proton colliders.
We present an overview in Section \ref{sec:overview}, and describe the tools we use and our general strategy in Section \ref{sec:strategy}.
Finally, in Section \ref{sec:results} we present a detailed description of the analyses performed, and a discussion of their results.

\subsection{Overview}
\label{sec:overview}

Pair produced Dark Matter particles can be searched for in events with large missing transverse energy (\met) in association with hard SM radiation.
The channels typically considered are the so called ``mono-X'' ones, where X can be a highly energetic jet or photon, but also a W boson, a Z boson, a Higgs etc. Within this category, we will focus on the monojet and monophoton channels. 
An additional strategy to look for Dark Matter in association with large \met\;is via vector boson fusion (VBF) processes. They are characterized by two forward jets with large invariant mass: these peculiar kinematical properties can be exploited to reduce the SM background and increase the sensitivity to Dark Matter particles with electroweak couplings. 
We will include this channel in our analysis.

In the searches mentioned so far, the signal receives contribution not only from  the neutral component of the electroweak multiplet, i.e. the Dark Matter candidate, but also from the electrically charged partner. Indeed, for the small mass splitting under consideration ($164\div165$ MeV), the charged component $\chi^{\pm}$ decays into $\chi_0$ and low-momentum charged pions $\pi^{\pm}$, which are not reconstructed at the {\sc Lhc}. Moreover its lifetime, $\tau \simeq 0.2$ ns, corresponds to a decay length at rest $d_0 = c \tau \simeq 6$ cm. Current detectors typically do not reconstruct charged tracks shorter than $\mathcal{O} (30)$ cm, therefore the bulk of the $\chi^{\pm}$ produced in partonic collisions contributes to the missing transverse momentum and energy of the events, in the same way as $\chi_0$.

Still, a small but non-negligible fraction of the $\chi^{\pm}$s, corresponding to the tail of the decay distribution, can travel enough to leave a track in the detector. These events would appear as high $p_T$ charged tracks ending inside the detector, once the $\chi^{\pm}$ decays into $\chi_0$ and soft undetected pions. At {\sc Lhc} with $\sqrt{s}=8$ TeV, searches of ``disappearing tracks'' provide the most sensitive probe of the scenario under consideration. The analysis presented by {\sc Atlas} in~\cite{Aad:2013yna} excludes $M_{\chi^{\pm}} <270$ GeV (95$\%$ CL) with $L=20.3$ fb$^{-1}$.
In this work we will study the expected sensitivity for disappearing tracks searches at future $pp$ colliders.

\medskip

Summarizing, we consider four possible channels to search for a stable fermion electroweak triplet:
monojet, monophotons, vector boson fusion and disappearing tracks. We compute the exclusion sensitivies for 
\begin{itemize}
\item the {\sc Lhc} with a center of mass energy of $\sqrt{s}=14$ TeV and an integrated luminosity of $L=3$ ab$^{-1}$, which is expected to be delivered in the High Luminosity (HL) run,
\item a futuristic $pp$ collider operating at $\sqrt{s}=$100 TeV, for $L=3$ ab$^{-1}$ and $L=30$ ab$^{-1}$.
The latter benchmark of integrated luminosity has for instance already been considered in \cite{Cohen:2014hxa} for stop searches.
\end{itemize} 

\subsection{Strategy}
\label{sec:strategy}
To perform our analysis we implement the model described by Eq.\eqref{eq:Lagrangian} in FeynRules 2.0~\cite{Alloul:2013bka}.
The events are simulated using MadGraph5~\cite{Alwall:2011uj} at the matrix element level and Pythia 6.4~\cite{Sjostrand:2006za}\footnote{We remark that Pythia does not (yet) include the effects of EW radiation, which might be important for a collider operating at  $\sqrt{s} = 100$ TeV, as recently pointed out in\cite{Hook:2014rka}.}, included in the default MadGraph package, for showering and hadronization. We use the `cteq6l1' pdf set. We adopt Delphes 3 as a detector simulator~\cite{deFavereau:2013fsa}, using for definiteness the default $\mbox{delphes}\_\mbox{Cms}$ card \footnote{Some of the analyses have been repeated using the PGS detector simulator~\cite{pgs}. We find similar results.}.

\medskip
For each channel we simulate the most relevant SM backgrounds and the signal. Then, we investigate the most appropriate selection cuts on the kinematical variables in order to maximize the significance, which is computed as follows:

\begin{equation}
\mbox{Significance}= \frac{S}{\sqrt{B + \alpha^2 B^2 + \beta^2 S^2}}\,,
\label{eq:significance}
\end{equation}
where $S$ and $B$ are respectively the expected number of signal and background events passing the cuts.
In Eq.\eqref{eq:significance} we sum in quadrature the statistic and systematic uncertainties. We denote the latter ones as $\alpha$ for the background and $\beta$ for the signal. We then consider two possible scenarios: an optimistic one, corresponding to $\alpha = 1\%$, and a more conservative one, where we fix $\alpha = 5\%$. For the monojet and monophoton analyses, this second value is in line with the systematics quoted by current experimental searches, see~\cite{CMS-PAS-EXO-12-048,ATLAS-CONF-2012-147} and~\cite{CMS-PAS-EXO-12-047,Aad:2012fw} respectively. For the VBF analysis it instead corresponds to a moderate improvement with respect to the present situation, which we infer from invisible Higgs decay searches~\cite{Chatrchyan:2014tja}, given that analogous DM searches in VBF have not yet been published. For simplicity of exposition, we stick to the same value of 5\% also for this last analysis. 
The disappearing tracks channel deserves special scrutiny, and we refer to Sect. \ref{Disappearing} for a discussion of the systematic uncertainties associated with the backgrounds.
Concerning the signals, we assume $\beta=10 \%$ for all the analyses under consideration, and this has only a marginal impact on our results.

\subsection{Analyses and results}
\label{sec:results}

\subsubsection{Monojet}
\label{Monojet}

Monojet searches require a hard central jet and large \met, and they have been performed at the 8 TeV {\sc Lhc} by the {\sc Atlas} and {\sc Cms} collaborations~\cite{CMS-PAS-EXO-12-048,ATLAS-CONF-2012-147}.

\begin{figure}[t]
\centering
\includegraphics[width=0.327\textwidth]{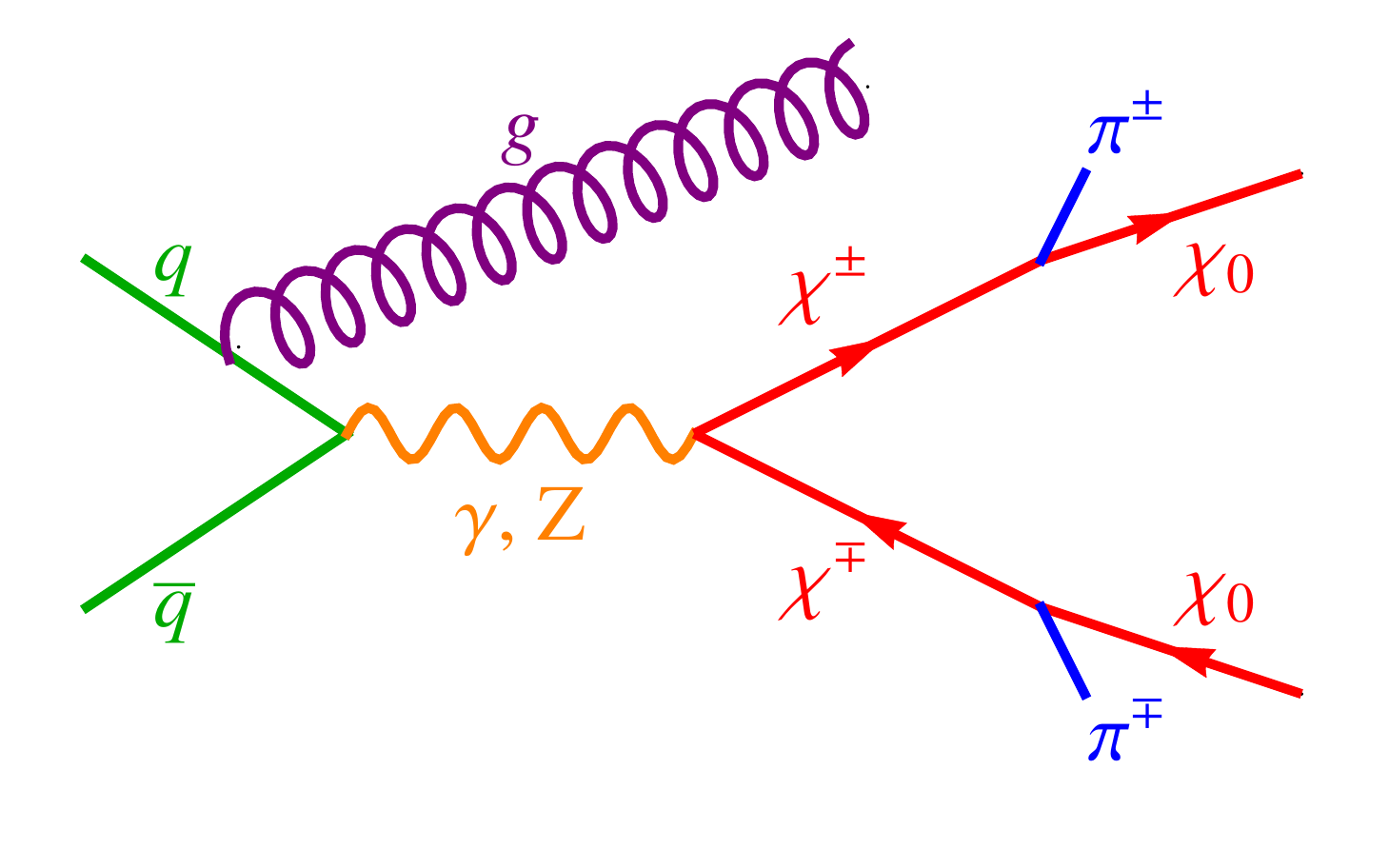}
\includegraphics[width=0.327\textwidth]{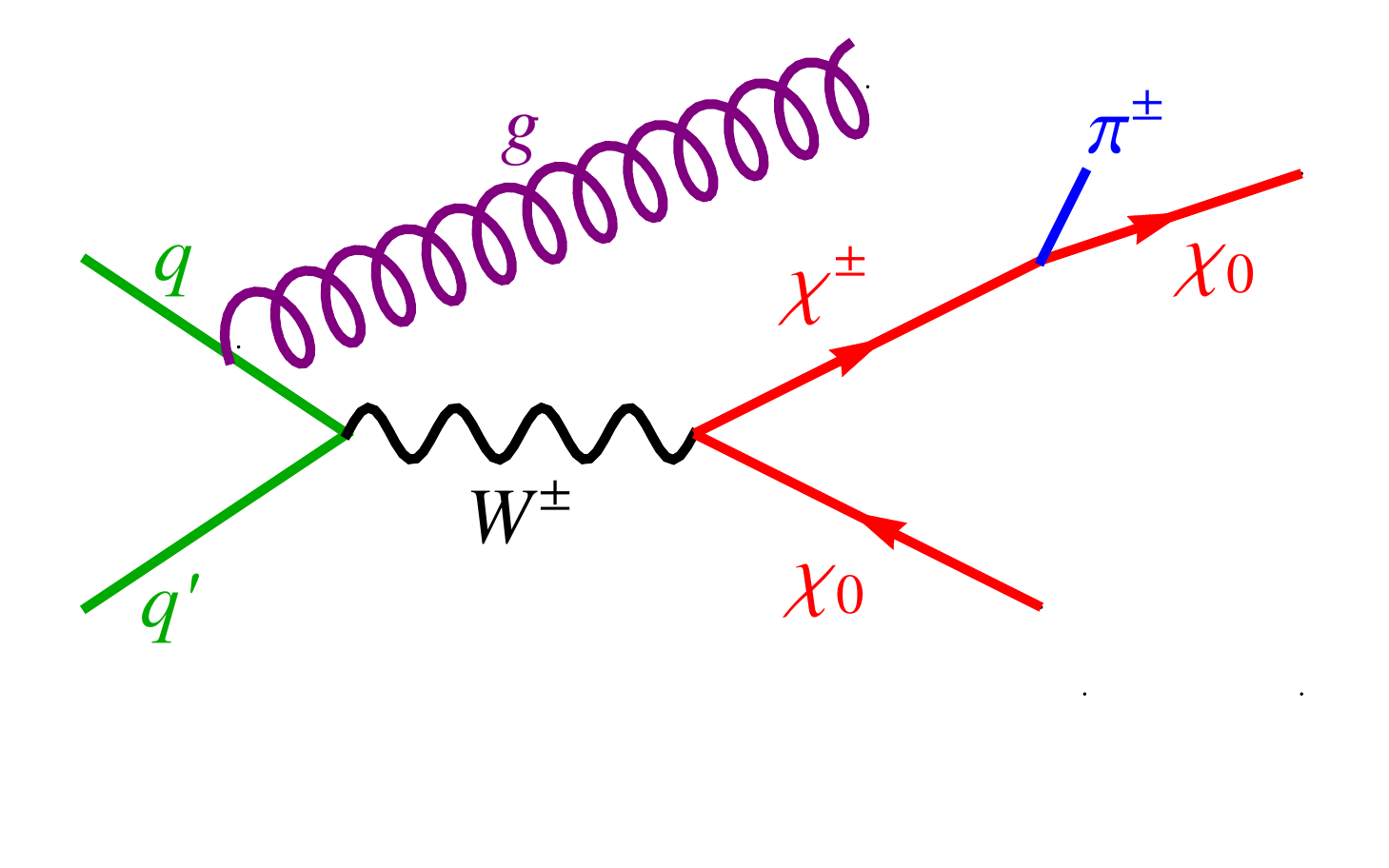}
\includegraphics[width=0.327\textwidth]{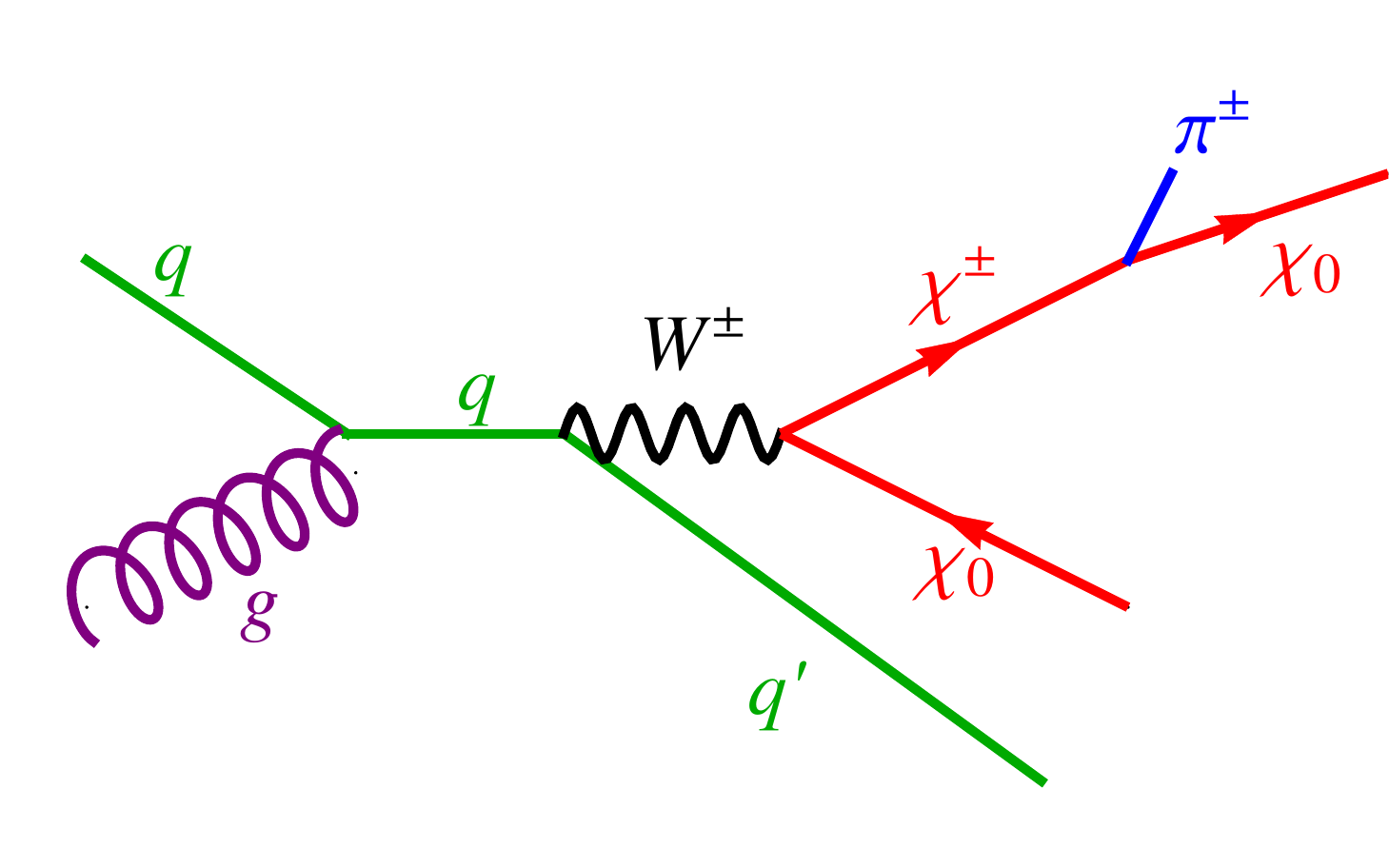}
\caption{Illustration of some Feynman diagrams for monojet processes.}
\label{Fig:diagrams_monoj}
\end{figure}

\smallskip
The signal is produced by processes like those in Fig.~\ref{Fig:diagrams_monoj}.
The dominant backgrounds are Z+jets with the Z boson decaying into neutrinos, and W+jets with the W decaying leptonically and the lepton is either undetected (too soft or close to the beam axis) or mistagged. Further background processes, which in~\cite{CMS-PAS-EXO-12-048,ATLAS-CONF-2012-147} account for less than $2 \%$ of the total event rate, are: $\mbox{t}\bar{\mbox{t}},$ Z($\ell \ell)$+jets, single t and QCD multijets. We discard these subdominant backgrounds from our analysis.

\smallskip
We first validate our simulation, matching one and two jets samples, against the analysis of Ref.~\cite{CMS-PAS-EXO-12-048} performed at 8 TeV with $L = 19.6$ fb$^{-1}.$ We find a good agreement in the expected number of Z($\nu\bar{\nu}$)+jets and W($\ell \nu$)+jets background events, after applying the analysis cuts.

\begin{table}[!t]
\vspace{.5 cm}
\centering
\renewcommand{\arraystretch}{1.2}
\renewcommand\tabcolsep{5pt}
\begin{tabular}{
|r|ccc|} \hline
      Cuts& 14 TeV & 100 TeV 3 ab$^{-1}$ & 100 TeV 30 ab$^{-1}$ \\ \hline
	\met [TeV]& $0.8 -1.6$ & $3 - 7$ & $3 - 7$\\ 
       $p_T(j_1)$  [TeV]  &0.4 &  1.4 & 1.5 \\ 
       $p_T(j_2)$ [GeV]  & $50 - 250$ & $100 - 500$  & $100 - 500$ \\  
       $\eta_1$    &  2.2 & 2.2 & 2.2 \\  
       $\Delta \phi$    &  2.2 & 2.2 & 2.2 \\  
       $p_T(\ell)$ [GeV]   &  20 & 20 & 20\\  
       $p_T(\tau)$ [GeV]    &  30 & 40 & 40 \\  \hline
 \end{tabular}\vspace{0.3cm}
\caption{\label{tab:monojet} Analysis cuts for the monojet search at 14 TeV and 100 TeV colliders.}
\end{table}

We simulate the backgrounds and the signal at 14 TeV and 100 TeV and, following the available experimental searches, we impose the following cuts:

\begin{itemize}

\item[$\circ$] we require missing transverse energy $>$ \met ,

\item[$\circ$]  we require the jet to be hard, i.e. with transverse momentum $p_T>p_T(j_1),$  and central, i.e. with pseudorapidity $\eta< \eta_1$,

\item[$\circ$]  a second jet with $p_T>p_T(j_2)$, $|\eta | < 4.5$, and azimuthal separation from the leading jet $< \Delta \phi$ is allowed, while additional jets are vetoed,

\item[$\circ$]  events with leptons are vetoed if the lepton has $\eta < 2.5$ and $p_T > p_T(\ell)$ (electrons and muons), $p_T > p_T(\tau)$ (taus).

\end{itemize}

The analysis cuts are summarized in Table~\ref{tab:monojet}. Two of them, \met \;and $p_{T}(j_2)$, are left free to vary over the ranges specified in the table, while the others are fixed. For each mass $M_{\chi_0}$, we compute the sensitivities over the 2-D grid of \met \;-$pt(j_2)$ cuts, and then we select the largest one. We note that the best choice of the analysis cuts depends on the assumption for the systematic uncertainties of the background (i.e. 1\% or 5\%). In particular larger systematics typically demand tighter cuts, as expectable from the way the significance scale with the number of events \eqref{eq:significance}, different in the two cases of systematics- and statistics- domination.
The values of cuts which are kept fixed have been previously determined in a more complete scan, where they have all been left varying. A priori, the precise choice of those cuts that maximizes the sensitivity depends on the mass $M_{\chi_0}$ of the simulated signal. However we find that fixing those cuts, for all the masses, to the values reported in Table~\ref{tab:monojet} has a small impact on the final sensitivity. Therefore, for simplicity we fix their values for all the signal masses.

\begin{figure}[t]
\centering
\includegraphics[width=10 cm]{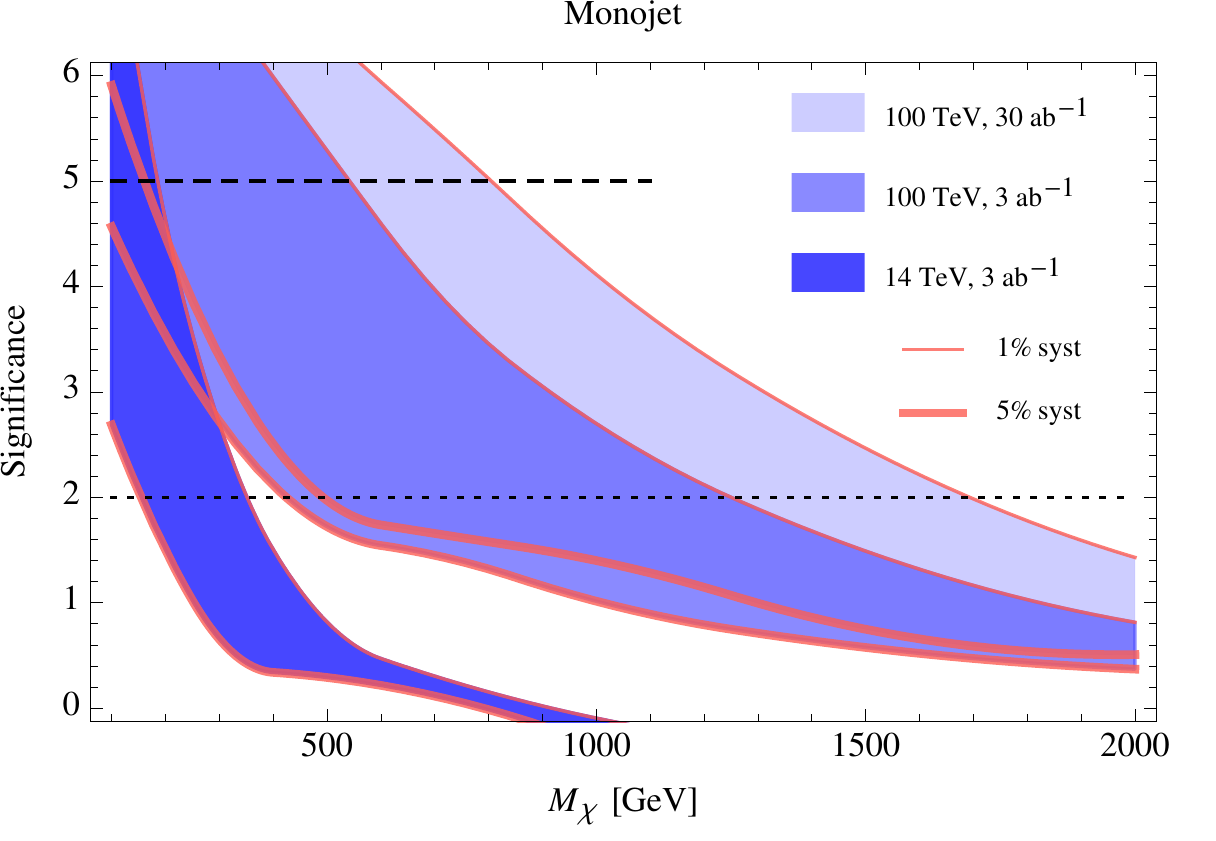}
\caption{Reach of monojet searches.}
\label{Fig:monojet}
\end{figure}

\medskip 

The results are shown in Fig.~\ref{Fig:monojet}. The 95 \% CL reach of {\sc Lhc-14} with L~=~3~ab$^{-1}$ is at the level of $M_{\chi}\sim$ 350 or 150 GeV, depending on the choice of systematic uncertainty of the background (as previously discussed, we fix either 1\% or 5\%). We find that a 100 TeV collider can improve the reach of a factor 3-4 with respect to {\sc Lhc-14}.  Systematic uncertainties play an important role in the determination of the sensitivity, especially at a 100 TeV collider. In particular raising the luminosity to L~=~30~ab$^{-1}$ would produce only a modest improvement of the sensitivity, for a systematic uncertainty of $\alpha=5$\%. However, it is not implausible that for such a high luminosity a better control of systematic uncertainties will be achieved.\\
Our findings are in good agreement with those of Ref.~\cite{Low:2014cba}, where the monojet reach has been quantified for 14 and 100 TeV $pp$ colliders with L~=~3~ab$^{-1}$.

\subsubsection{Monophoton}
\label{monophoton}

Monophoton searches at the {\sc Lhc} have been performed by the {\sc Atlas} and {\sc Cms} collaborations~\cite{CMS-PAS-EXO-12-047,Aad:2012fw}.
These analyses require a high $p_T$ photon in addition to large \met. Quality criteria and isolation requirements are imposed to the photon candidate.

\begin{figure}[t]
\centering
\includegraphics[width=0.327\textwidth]{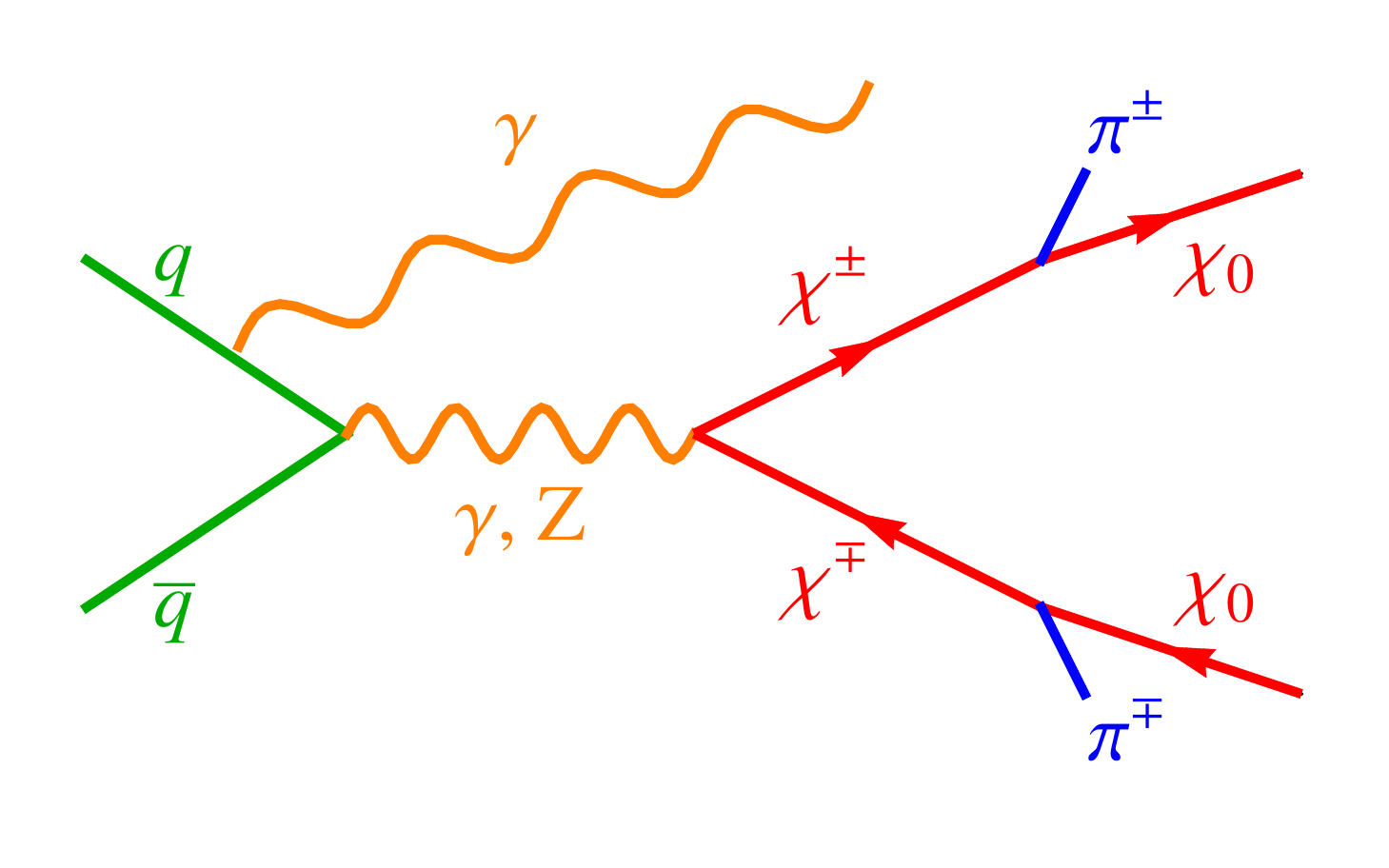}
\includegraphics[width=0.327\textwidth]{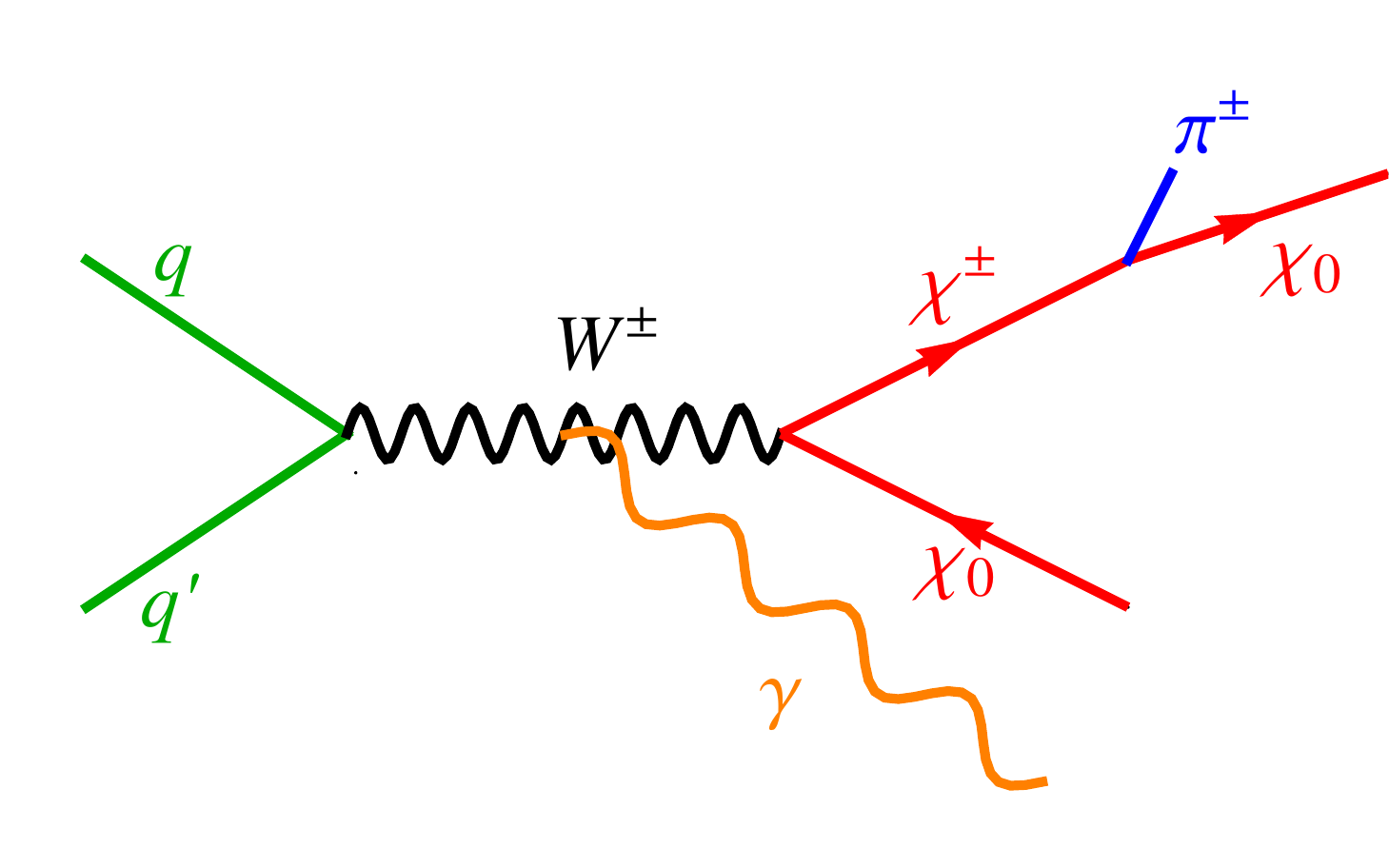}
\includegraphics[width=0.327\textwidth]{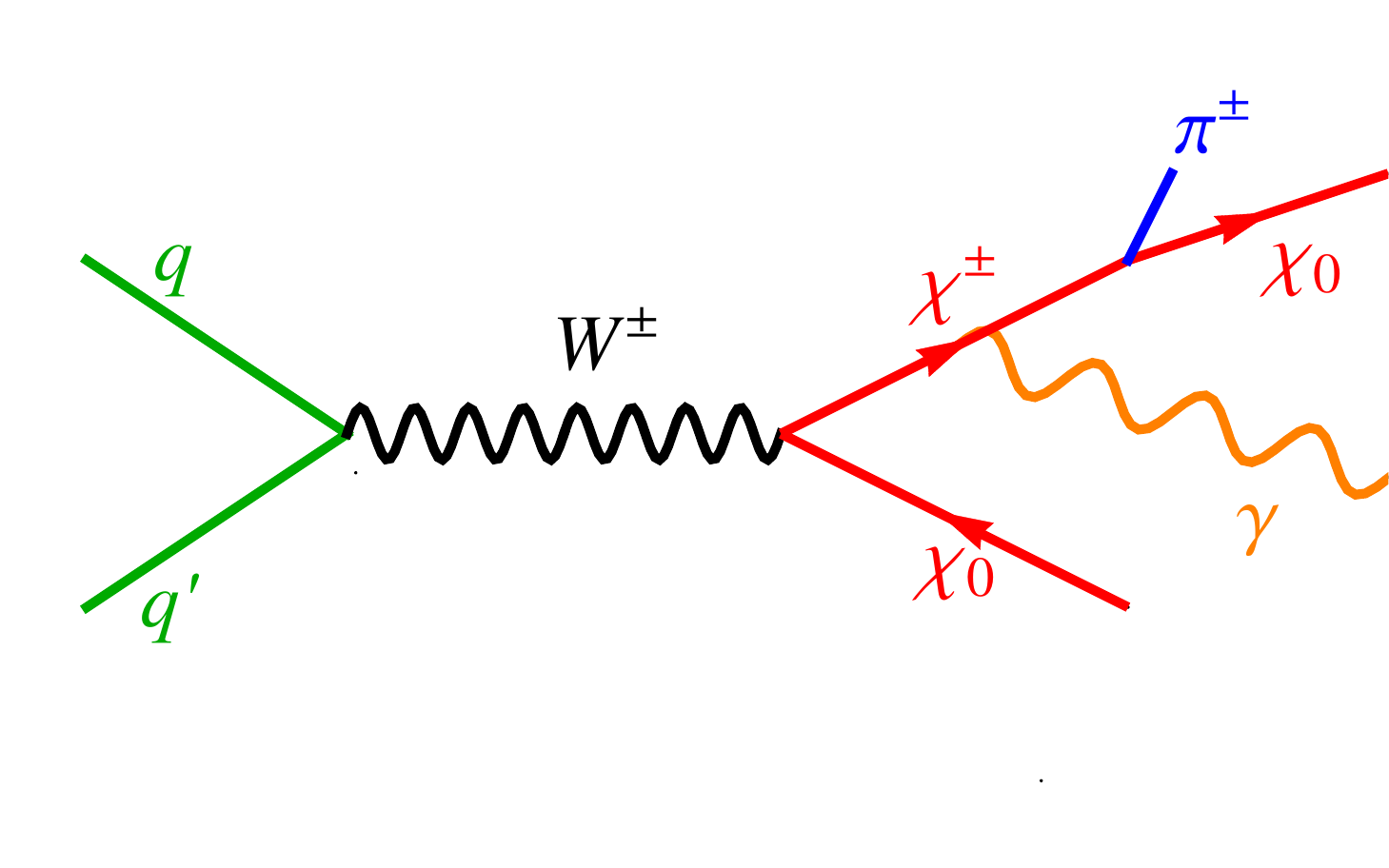}
\caption{Illustration of some Feynman diagrams for monophoton processes.}
\label{Fig:diagrams_monophoton}
\end{figure}

The largest background comes from $\gamma Z(\bar{\nu} \nu)$ processes. Additional backgrounds include $\gamma W(\ell \nu)$, $W(\ell \nu)$, $\gamma$+jets, multijet, $\gamma Z(\ell \ell)$ and diphoton.
Signal processes are for instance those shown in Fig.~\ref{Fig:diagrams_monophoton}. Notice that a photon can also be radiated from the final state, as opposite to the cases where the hard SM radiation on which one tags is constituted of jets, and also to other DM candidates where charged states do not contribute to the signal.
\smallskip

\begin{table}[!t]
\vspace{.5 cm}
\centering
\renewcommand{\arraystretch}{1.2}
\renewcommand\tabcolsep{5pt}
\begin{tabular}{
|r|ccc|} \hline
      Cuts& 14 TeV & 100 TeV 3 ab$^{-1}$ & 100 TeV 30 ab$^{-1}$ \\ \hline
	\met [TeV]& $0.3 -1$ & $1 - 3$  & $1 - 3.5$\\ 
       $p_T(\gamma)$  [GeV]  & $200 - 500$ & $500 - 700$  & $500 - 700$ \\ 
       $\eta_{\gamma}$    &  1.45 & 1.45 & 1.45 \\  
       $\Delta \phi$  &  2 & 2 & 2 \\  
       $p_T(j)$ [GeV]  & 30 & 100  &  100 \\  
       $p_T(\ell)$ [GeV]   &  20 & 20 & 20\\  
       $p_T(\tau)$ [GeV]    &  30 & 40 & 40 \\  \hline
 \end{tabular}\vspace{0.3cm}
\caption{\label{tab:monophoton} Analysis cuts for the monophoton search at 14 TeV and 100 TeV colliders.}
\end{table}

We first compare our procedure, which includes matching with the case of one extra jet, with the {\sc Cms} analysis at 8 TeV with $L =~19.6$ fb$^{-1}$~\cite{CMS-PAS-EXO-12-047}. The cuts that we implement are listed below (the precise values that we choose are those of~\cite{CMS-PAS-EXO-12-047}).
While we find good agreement for the background $\gamma W(\ell \nu)$, our estimate for the $\gamma Z(\bar{\nu} \nu)$ one is a factor 1.35 larger than that in~\cite{CMS-PAS-EXO-12-047}. This could be due to the fact
that we are missing some selection cuts on the photon that are particularly difficult to implement in our analysis. Similar results have been found in the phenomenological studies~\cite{Fox:2011pm,Hagiwara:2012we}.

For the projections at 14 TeV and 100 TeV colliders, we compute the background events including only $\gamma Z(\bar{\nu} \nu)$ and $\gamma W(\ell \nu)$ processes (which in the {\sc Cms} analysis at 8 TeV account for $\sim 75 \%$ of the total background events~\cite{CMS-PAS-EXO-12-047}). We therefore caution that some degree of uncertainty in the background estimation is present in our analysis. Still our computations should be a reasonable estimate of the potential reach of future hadron colliders with the monophoton search. 

\smallskip 

The analysis cuts that we impose are: 

\begin{itemize}

\item[$\circ$] we require missing transverse energy $>$ \met,

\item[$\circ$] we identify the leading photon as the one with the highest $p_T$ among those that have $p_T>p_T(\gamma)$  and pseudorapidity $|\eta|<\eta_{\gamma}$,

\item[$\circ$] the angular separation between the photon and \met \;should be larger than $\Delta \phi$,

\item[$\circ$] we discard events with more than one jet that has: i) $p_T>p_T(j)$, ii) $\eta<4.5$, iii) angular distance from the photon $\Delta R > 0.5$,

\item[$\circ$] events with leptons are vetoed if the lepton is $\Delta R>$0.5 away from the photon, and if it has $\eta < 2.5$ and $p_T > p_T(\ell)$ (electrons and muons), $p_T > p_T(\tau)$ (taus).

\end{itemize}

\begin{figure}[t]
\centering
\includegraphics[width=10 cm]{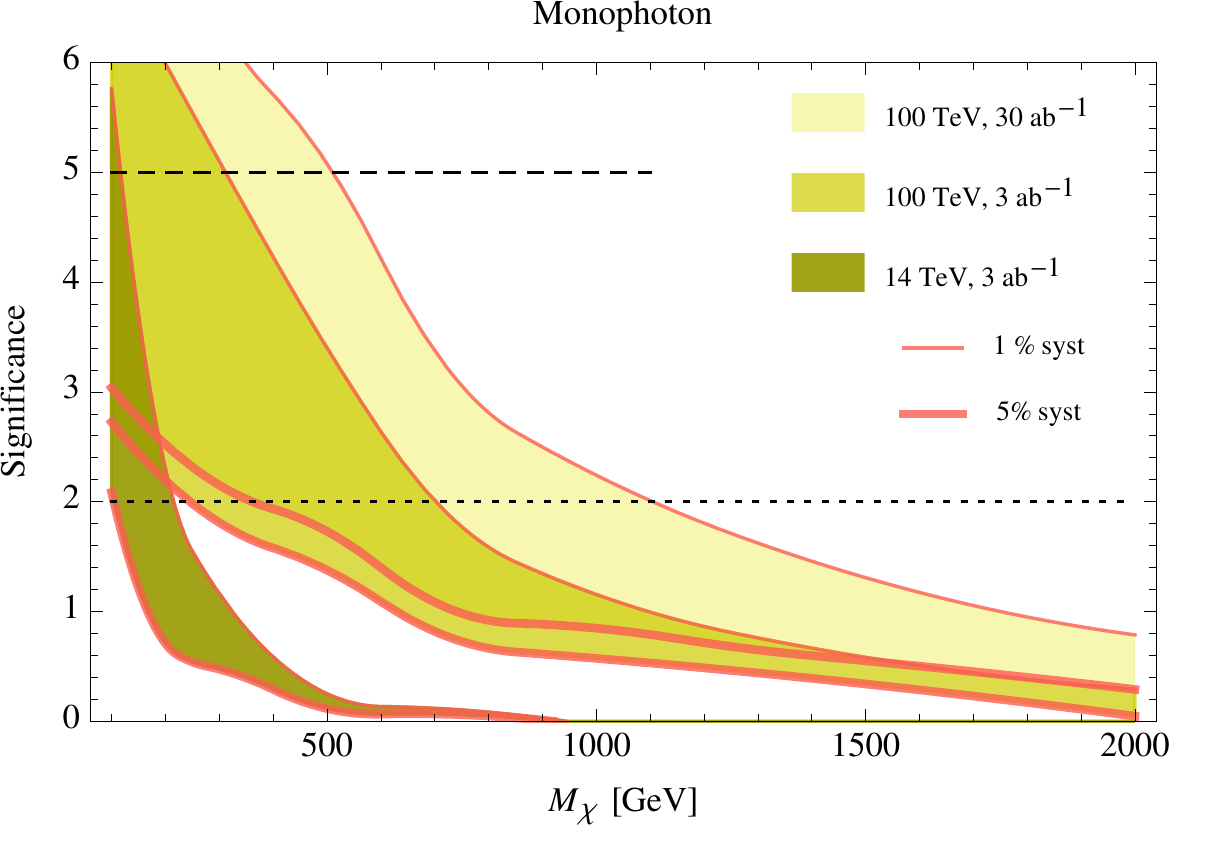}
\caption{Reach of monophoton searches.}
\label{Fig:monophoton}
\end{figure}
Table~\ref{tab:monophoton} summarizes the values of the cuts chosen. The $p_T$ of the photon and \met \;are scanned over the ranges specified in Table~\ref{tab:monophoton}. For each Dark Matter mass we compute the sensitivity for the points of the 2-D grid of cuts, and we select the maximal value. As for the monojet case, the values we kept fixed were preliminary determined scanning on a higher dimensional grid, which included $\eta_\gamma$, $\Delta \phi$ and $p_T(j)$.

\smallskip
Fig.~\ref{Fig:monophoton} shows our results. A 14 TeV collider will reach a 95 \% CL sensitivity for Dark Matter masses at the level of 200 or 100 GeV, depending on the choice of the systematic uncertainties (we recall that for the background systematics we choose either 1 or 5 \%).
We find that the reach at 100 TeV with L~=~3 ab$^{-1}$ will extend by a factor of 3-4 in mass, and that again a control of systematic uncertainties will play a crucial role in exploiting the potential of possible higher integrated luminosities. Among the searches that we analyse, the monophoton one turns out to be that with the lowest mass reach.

\subsubsection{Vector boson fusion}
\label{VBF}

Vector boson fusion processes have been investigated by the {\sc Cms} collaboration at {\sc Lhc-8} in order to search for invisible decay channels of the Higgs boson~\cite{Chatrchyan:2014tja}.
This channel can be exploited also to look for Dark Matter particles with electroweak interactions, like the candidate we are considering.
VBF processes are characterized by two forward jets in opposite hemispheres (i.e. well separated in pseudorapidity), and with a large invariant mass. Cuts on these variables as well as the requirement of large \met\;are used in order to reduce the SM background.

\begin{figure}[t]
\centering
\includegraphics[width=0.327\textwidth]{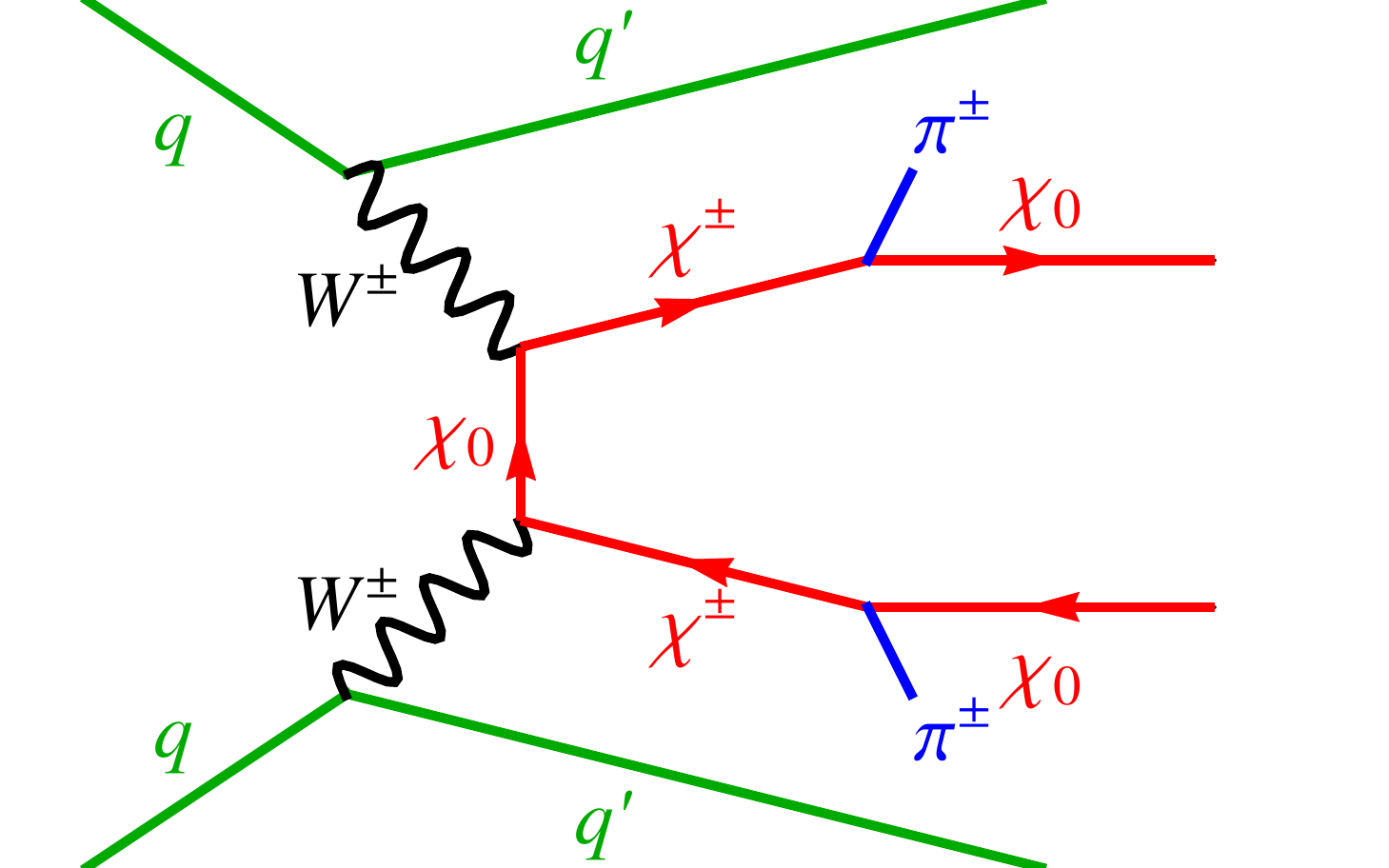}
\includegraphics[width=0.327\textwidth]{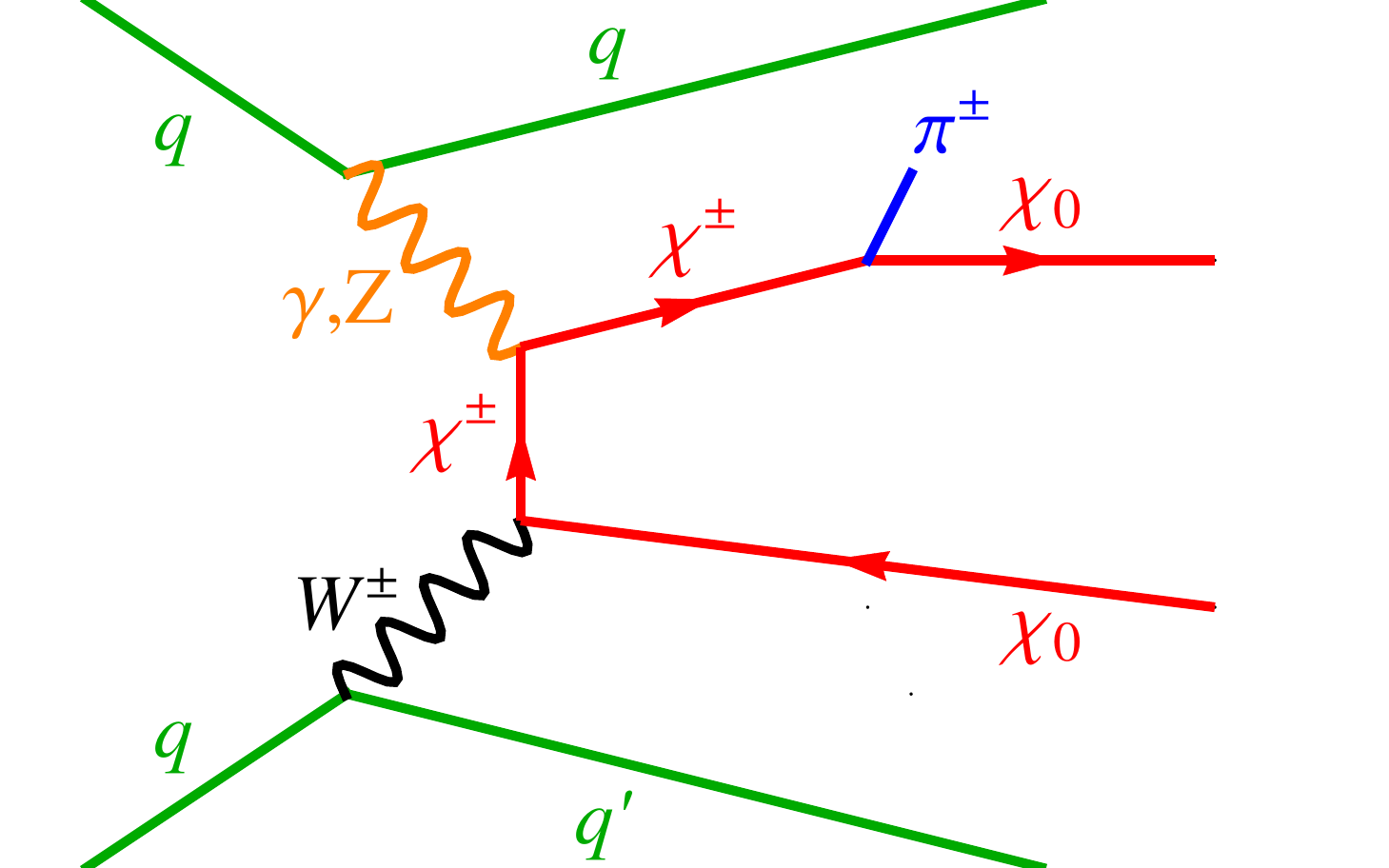}
\includegraphics[width=0.327\textwidth]{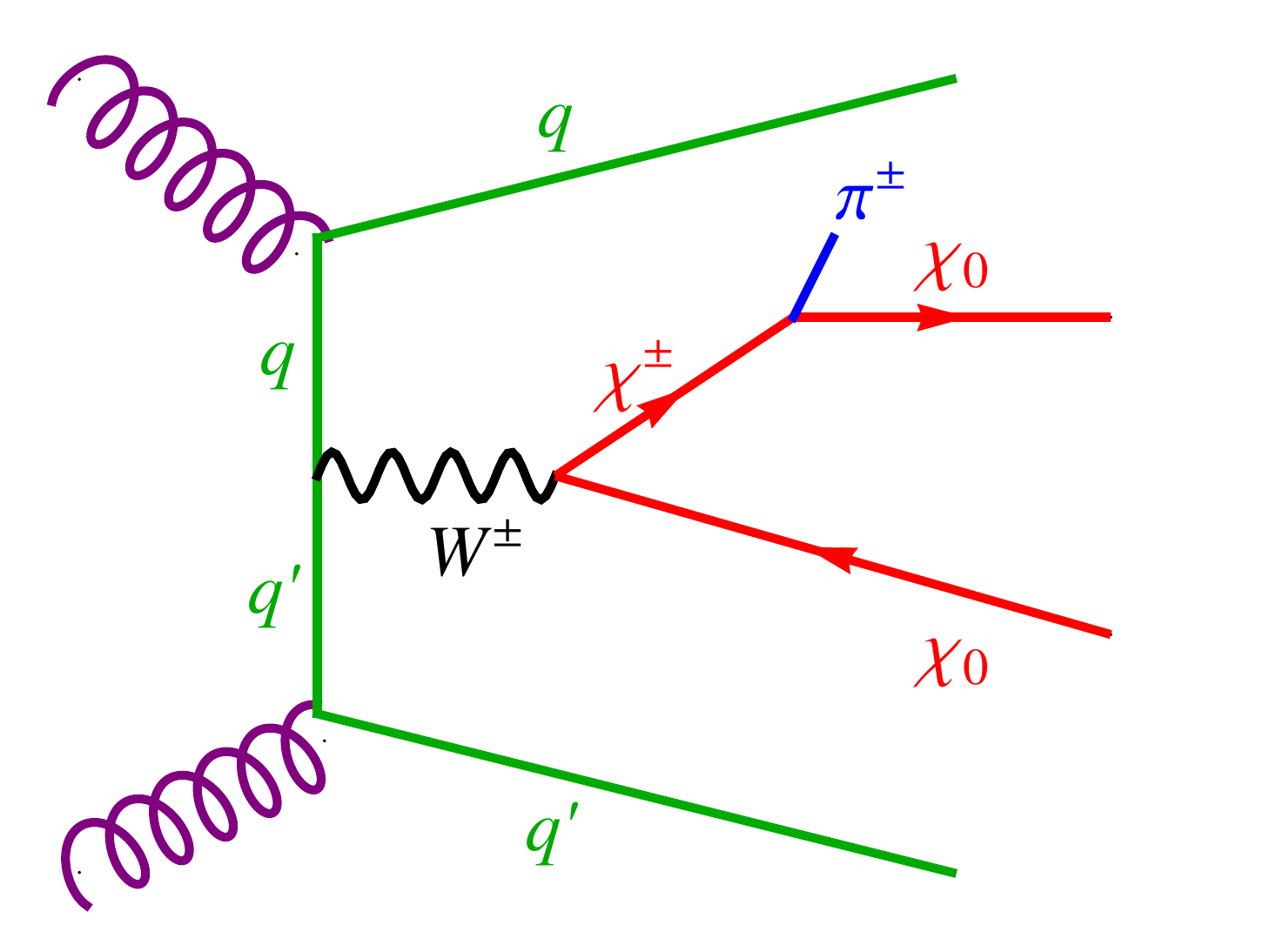}\\ [5 mm]
\includegraphics[width=0.327\textwidth]{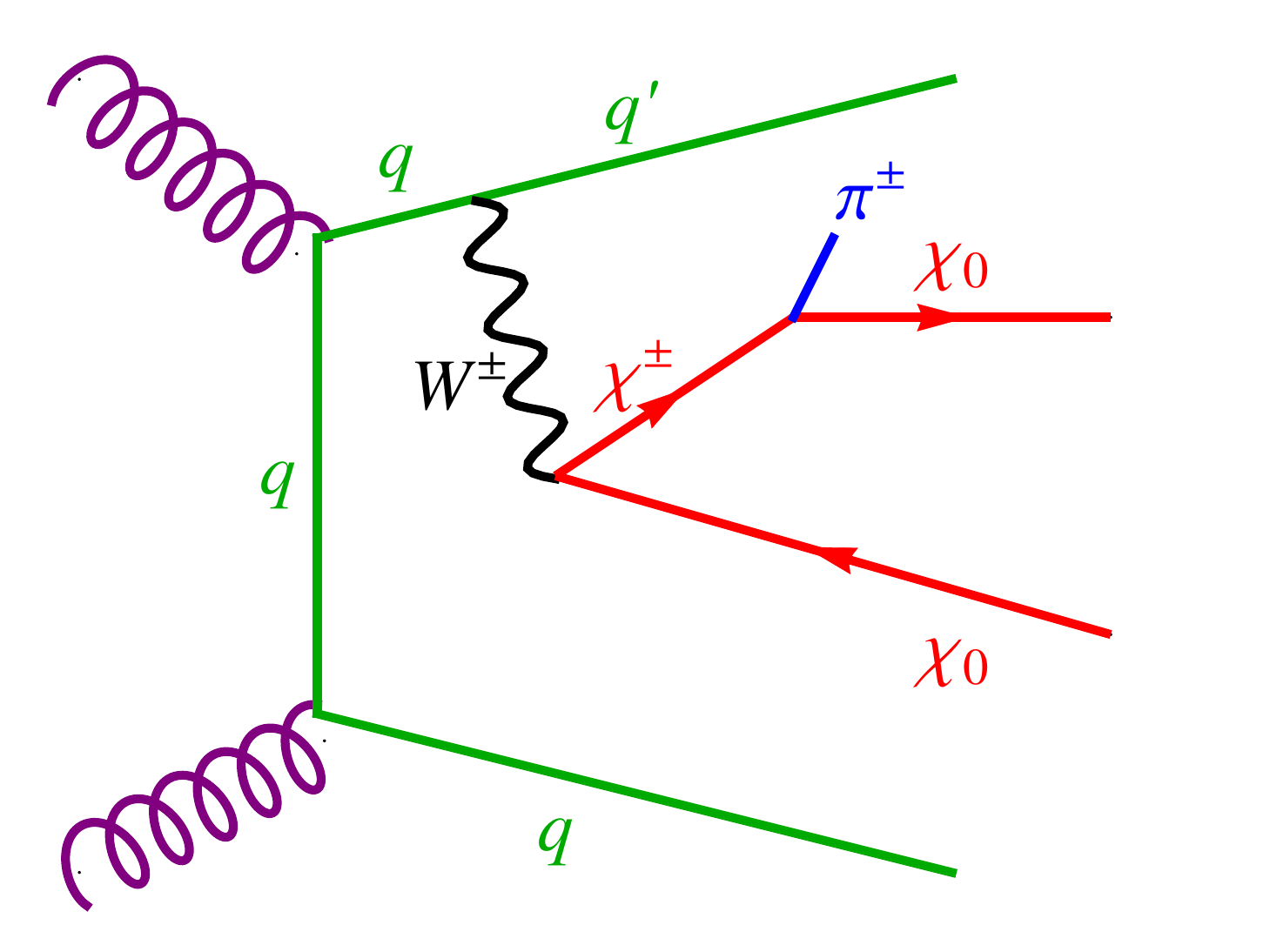}
\includegraphics[width=0.42\textwidth]{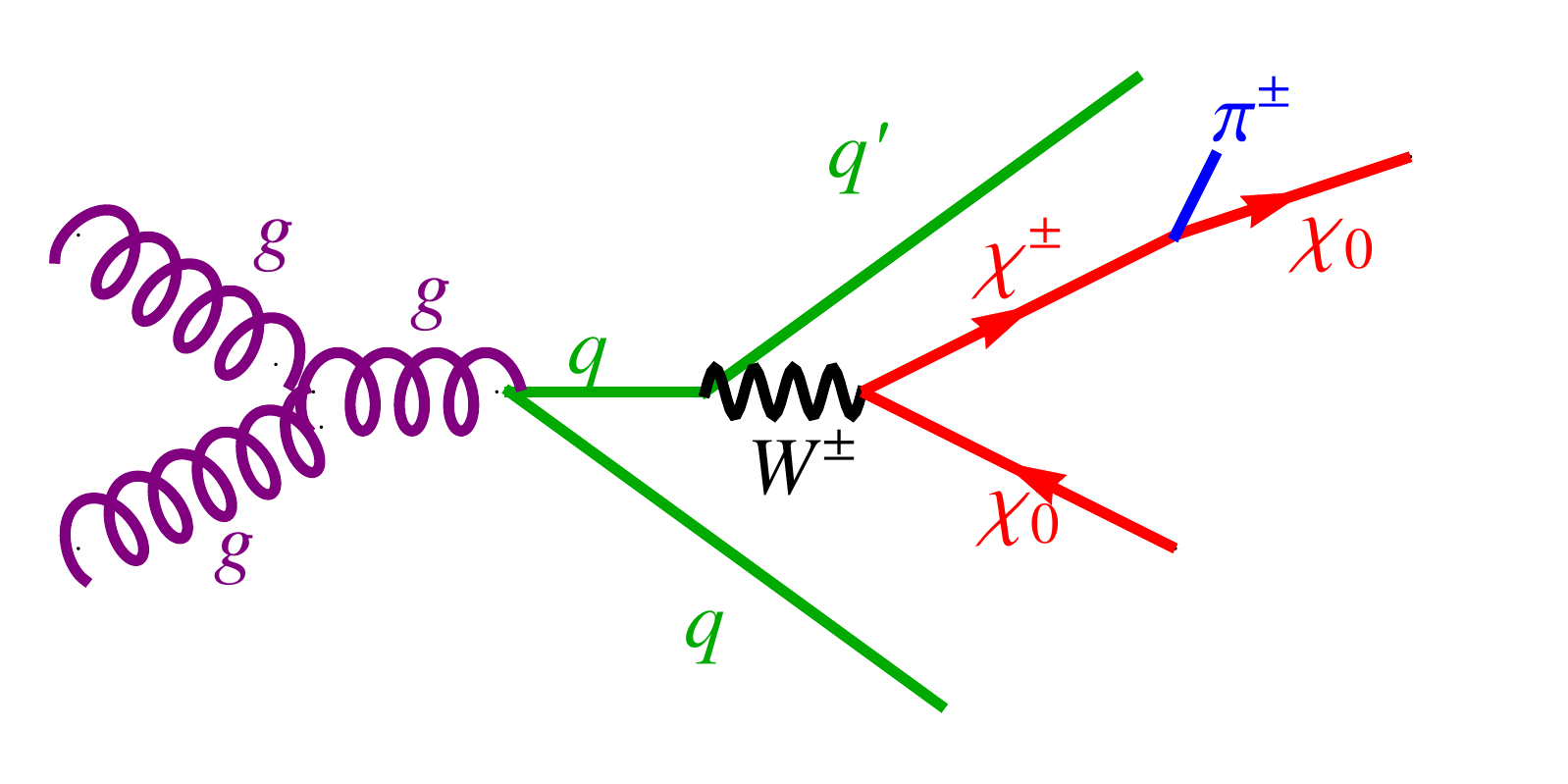}
\caption{Illustration of some Feynman diagrams for VBF processes.}
\label{Fig:diagrams_VBF}
\end{figure}

Examples of diagrams relevant for this search\footnote{Notice that, despite the conventional name of the channel, also diagrams not properly originating from two vector bosons contribute to the signal (and also background) events.} are shown in Fig.~\ref{Fig:diagrams_VBF}.
The dominant backgrounds result from Z($\nu\bar{\nu}$)+jets and W($\ell \nu$)+jets (where the lepton is lost) events. For example in the search of {\sc Cms} at 8 TeV~\cite{Chatrchyan:2014tja} they constitute $\sim$ 85\% of the total.

\begin{table}[!t]
\vspace{.5 cm}
\centering
\renewcommand{\arraystretch}{1.2}
\renewcommand\tabcolsep{5pt}
\begin{tabular}{|r|ccc|} \hline
      Cuts& 14 TeV & 100 TeV 3 ab$^{-1}$ & 100 TeV 30 ab$^{-1}$ \\  \hline
	\met [TeV]& $0.4-0.7$ & $1.5 - 5.5$ & $1.5 - 5.5$ \\ 
       $p_T(j_{12})$  [GeV] &  40 (1\%),\, 60 (5\%) & 150 & 200\\ 
       $M_{jj}$ [TeV]   &  1.5 (1\%),\, 1.6 (5\%) & 6 (1\%),\, 7 (5\%) & 7\\ 
       $\Delta \eta_{12}$  &  3.6 & 3.6 & 3.6 (1\%),\, 4 (5\%)\\
       $\Delta \phi$ & $1.5 - 3$ &  $1.5 - 3$ &  $1.5 - 3$\\
       $p_T(j_3)$  [GeV]    &  25 & 60 & 60\\  
       $p_T(\ell)$ [GeV]   &  20 & 20 & 20\\  
       $p_T(\tau)$ [GeV]    &  30 & 40 & 40 \\  \hline 
 \end{tabular}\vspace{0.3cm}
\caption{\label{tab:VBF} Analysis cuts for the VBF search at 14 TeV and 100 TeV colliders.}
\end{table}
\smallskip

We simulate the Z($\nu\bar{\nu}$)+jets and W($\ell \nu$)+jets backgrounds as well the signal for different Dark Matter masses at 14 and 100 TeV. As a check, we verify that we reproduce with good agreement the background counts of~\cite{Chatrchyan:2014tja}.
Like for the case of the monojet analysys, we first scan over several cuts on the kinematical variables, in order to optimize the sensitivity to the DM signal. We then identify the cuts which are more relevant to determine the sensitivity for different Dark Matter masses (which we find to be \met \;and, to a lower extent, tha azimuthal separation of the leading jets $\Delta \phi$), and for simplicity we fix the remaining ones.

The final analysis cuts are the following:

\begin{itemize}

\item[$\circ$]  we require missing transverse energy $>$ \met,

\item[$\circ$]  we require two leading jet, defined as those with the largest $p_T$, each of them satisfying $p_T>p_T(j_{12})$  and $|\eta|<4.5$ .

\item[$\circ$]  the two leading jets should also be well separated in pseudorapidity, $| \Delta \eta |>\Delta \eta_{12}$ and $\eta_1 \cdot \eta_2<$0

\item[$\circ$] they should have a high invariant mass $M_{j_1 j_2}> M_{jj}$,

\item[$\circ$] and their azimuthal separation should not exceed $\Delta \phi$,

\item[$\circ$]  we reject events with additional jets satisfing $p_T>p_T(j_3)$, $|\eta|< 4.5$ and pseudorapidity between the two tagged jets,

\item[$\circ$]  events with leptons are vetoed if the lepton has $\eta < 2.5$ and $p_T > p_T(\ell)$ (electrons and muons), $p_T > p_T(\tau)$ (taus).

\end{itemize}

\begin{figure}[t]
\centering
\includegraphics[width=10 cm]{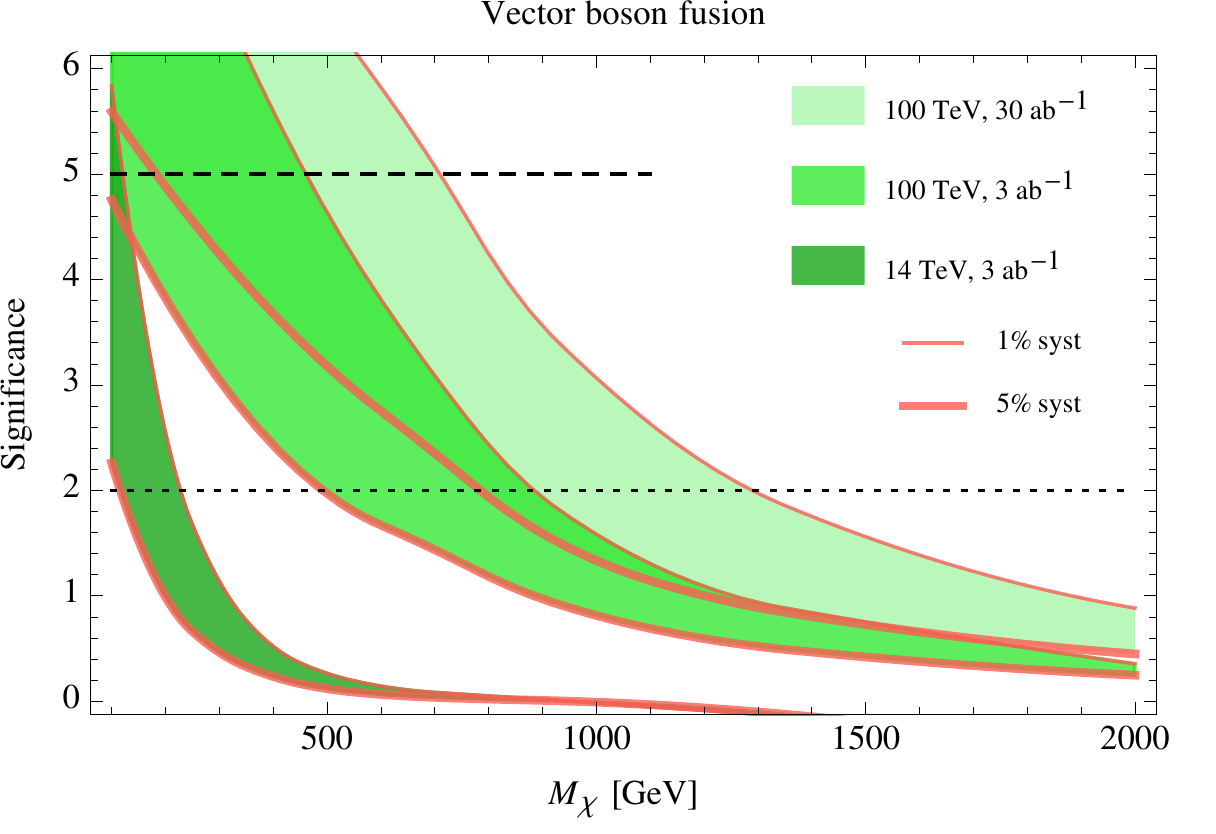}
\caption{Reach of VBF searches.}
\label{Fig:VBF}
\end{figure}

The analysis cuts are shown in Table~\ref{tab:VBF}. For each value of the Dark Matter mass we scan over the ranges of values of $\met$ \;and $\Delta\phi$ in Table~\ref{tab:VBF} and we identify the maximal sensitivity.

The results are shown in Fig.~\ref{Fig:VBF}. The 95 \% CL reach at a 14 TeV collider is at the level of 250 and 100 GeV, respectively for 1\% and 5\% systematics. For the same range of the systematics, the reach of a 100 TeV collider is found to lie between 500 and 900 GeV (L~=~3 ab$^{-1}$), and 800 and 1300 GeV (L~=~30 ab$^{-1}$)
The VBF search appears somehow less promising than the monojet one. Still, over a wide range of masses, a possible discovery in the monojet channel would also be confirmed with VBF processes. This will be a precious piece of information in order to constrain the properties of a possible future Dark Matter signal. 
We finally note that our expected sensitivities are significantly lower than those found in~\cite{Delannoy:2013ata}, where the reach of the VBF channel for Winos has been analysed at the {\sc Lhc-14}. While we have not been able to ultimately trace back the origin of this discrepancy, it can be useful to mention two of the several checks we performed: i) this difference is not simply ascribable to the fact that, in~\cite{Delannoy:2013ata}, systematics effects have been neglected, ii) we obtain a higher background count at high \met\;cuts.

\subsubsection{Disappearing Tracks}
\label{Disappearing}

Disappearing tracks signatures have received significant attention in the context of SUSY Winos~\cite{Feng:1994mq,Feng:1999fu,Gunion:1999jr,Gunion:2001fu,Barr:2002ex,Ibe:2006de,Buckley:2009kv,Kane:2012aa,Low:2014cba}.
As mentioned in Section~\ref{sec:collider}, searches at {\sc Lhc-8} exclude $M_{\chi^{\pm}} <270$ GeV at 95$\%$ CL.
The analysis has been performed by the {\sc Atlas} collaboration using 20.3 fb$^{-1}$ of data~\cite{Aad:2013yna}.
This search requires large \met, a jet with large $p_T$ to trigger the signal event, and at least a track with high $p_T.$
The candidate track should satisfy additional criteria, for instance to ensure well reconstruction and isolation.

The background originates from charged hadrons interacting with the inner detector, unidentified leptons (lepton tracks) and charged particles with highly mismeasured $p_T$. In the {\sc Atlas} analysis, the latter background is largely the dominant one for $p_T$ of the track ($x=p_T^{track}$) higher than 100 GeV, and it is found to be fitted by a power law $d\sigma/dx \propto x^{-a},$ with $a=1.78\pm 0.05.$
The expected background events are not estimated, in the experimental analysis, by means of MC simulations. Rather the $p_T$ shape for the different sources of background is identified using data in appropriate control regions. Then, to determine the reach of this search, the observed $p_T$ distribution of the tracks is fitted with the signal and background templates.

\smallskip

For these reasons, a precise determination of the sensitivity of this channel at future colliders looks particularly complicated. 
Following~\cite{Low:2014cba} we adopt instead a simple prescription. We assume the background will still be dominated by charged particles with highly mismeasured $p_T$, and we take the power law behaviour of the $p_T^{track}$ distribution previously mentioned. We fix the normalization by matching with the number of observed events in~\cite{Aad:2013yna}. Then, we assume that the bulk of the background is originated by $Z(\nu\bar{\nu})$+jets processes. We extrapolate the 8 TeV background at other center of mass energies in the following way: we extract the background cross section at 8 TeV, and we rescale it at higher energies with the ratio of the $Z(\nu\bar{\nu})$+jets cross-sections at those energies, computed with the appropriate analysis cuts (see below the cuts considered for 14 TeV and 100 TeV colliders and~\cite{Aad:2013yna} for those at 8 TeV). 
In order to account for the large uncertainties introduced with this procedure, the sensitivity is estimated also for two extreme cases, corresponding to a further multiplication of the background events by a factor of 5 and 1/5.
One could also estimate the uncertainty on the background events varying the index $a$ of the $p_T^{track}$ distribution. We checked that changing $a$ around $\pm 5 \sigma$ from its central value, produces a smaller uncertainty band than the method that we have adopted.

\smallskip

For the signal we perform our analysis simulating events with $\chi^{\pm}$ and a jet with MadGraph and Pythia, including matching with the case of one extra jet. For the sake of illustration, we show some channels for signal production in Fig.\ref{Fig:diagrams_monoj}. These processes are common also to monojet searches. However, we remind that in this analysis the $\chi^{\pm}$ is required to decay well inside the detector.
In our analysis, the $\chi^{\pm}$ decays are simulated with an exponential decay law, with the lifetime $\tau$ computed in~\cite{Ibe:2012sx}. The radial distance travelled by the track in the laboratory frame, $d$, is then $d=\beta \gamma c \tau.$
Finally, we apply the following cuts, which follow those in~\cite{Aad:2013yna}:
\begin{table}[!t]
\centering
\renewcommand{\arraystretch}{1.2}
\renewcommand\tabcolsep{5pt}
\begin{tabular}{|r|ccc|} \hline
      Cuts& 14 TeV & 100 TeV 3 ab$^{-1}$ & 100 TeV 30 ab$^{-1}$ \\ \hline
	\met \;[TeV]& 0.22 & 1.4 & 1.4 \\
       $p_T(j_{1})$ [TeV]  &  0.22  & 1.0 & 1.0  \\ 
       $\eta(j_{1})$  &  2.8 & 2.8 & 2.8\\ 
       $p_T(j_2)$ [GeV]  &  70 & 500 & 500\\  
       $p_T^{track}$ [TeV]    &  0.32 & 2.1  & 2.1\\ \hline
 \end{tabular}\vspace{0.3cm}
\caption{\label{tab:disappearing} Analysis cuts for the disappearing track search at 14 TeV and 100 TeV colliders.}
\end{table}

\begin{itemize}

\item[$\circ$]  we require missing transverse energy $>$ \met,

\item[$\circ$]  we require at least one jet with $p_T>p_T(j_{1})$ and $|\eta|<\eta(j_{1})$,

\item[$\circ$]  we compute the azimuthal separation between the leading jet and \met, $\Delta \phi^{j-\met}$.
	If the event contains multiple jets with $p_T>p_T(j_2)$, we consider also $\Delta \phi^{j-\met}$ of the second jet. The smallest $\Delta \phi^{j-\met}$ is then used. The event is required to have $\Delta \phi^{j-\met}>1.5$.

\item[$\circ$]  The event is required to contain at least one track with $p_T>p_T^{track}$ and $0.1<|\eta|<1.9$,

\item[$\circ$] the track should be isolated, therefore we reject events with jets residing in a cone of $\Delta R<0.4$ around the track and with $p_T>p_T(j_2),$

\item[$\circ$]  the track should have a radial length $30<d<80$ cm in order to be properly reconstructed by the tracker,

\item[$\circ$]  events with reconstructed electrons and muons are vetoed.

\end{itemize}

We first simulate the signal at 8 TeV with $L = 20.3$ fb$^{-1}$, and we apply the cuts of~\cite{Aad:2013yna}. The number of signal events is then matched to the value in~\cite{Aad:2013yna} multiplying for an efficiency $\epsilon,$ that we find to be $\epsilon=0.51$.
We use this efficiency also for the analysis at 14 and 100 TeV colliders. The analysis cuts we use are shown in Table~\ref{tab:disappearing}. We have determined them by scanning on a 3-D grid in \met\,, $p_T^{track}$ and $p_T(j_{1})$, and choosing those that gave a high sensitivity without reducing the event counts below the level of a few. The sensitivity is computed fixing $\alpha=20 \%$ and $\beta=10 \%$ in eq.~\ref{eq:significance}.

\begin{figure}[t]
\centering
\includegraphics[width=10 cm]{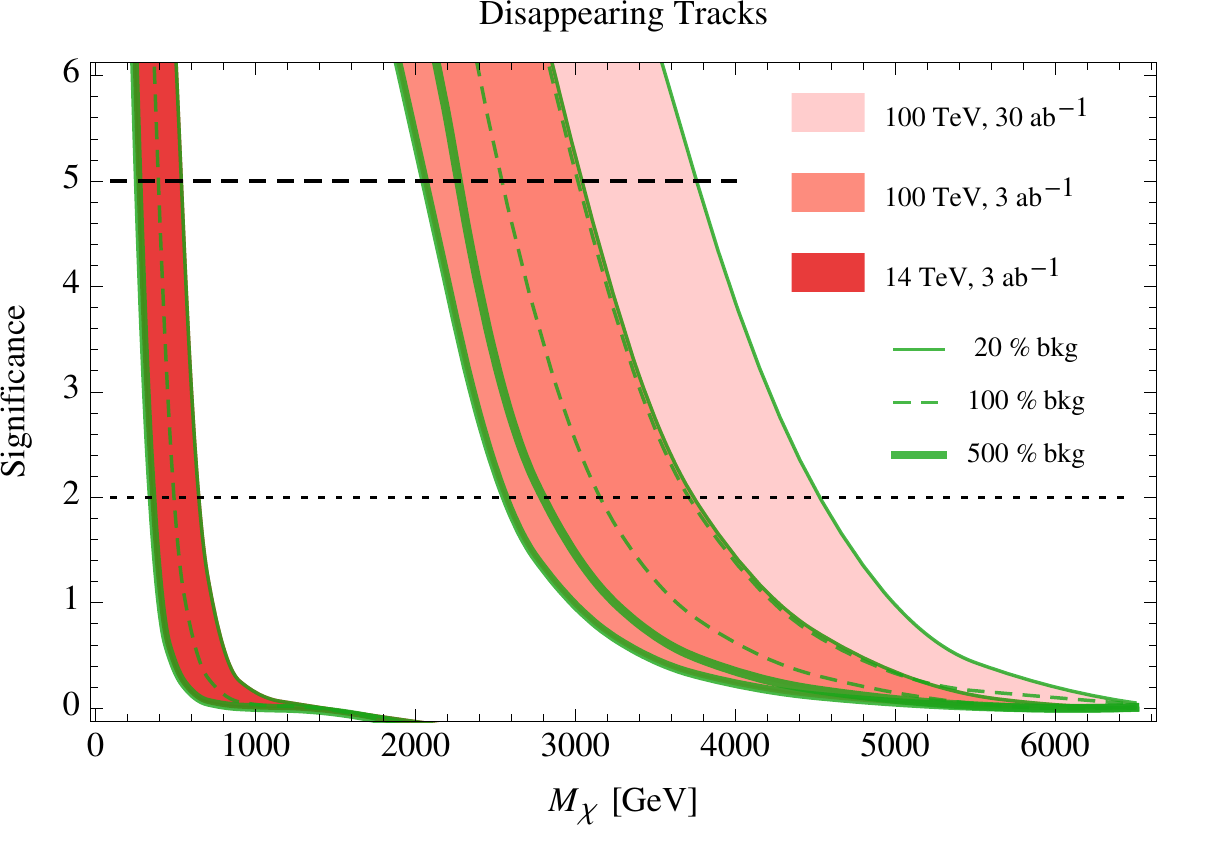}
\caption{Reach of disappearing tracks searches}
\label{Fig:DisTracks}
\end{figure}

The results are shown in Fig.~\ref{Fig:DisTracks} , with the band referring to the two choices of background estimation (i.e. the expected number of background events is multiplied or divided by a factor 5). In each band, we also show a dashed line corresponding to the central value for the background, i.e. in the absence of factors of 5. Among the searches we have considered, disappearing tracks are the most promising. At a 100 TeV collider they have a good chance to probe the thermal Dark Matter scenario, i.e. $M_{ \chi_{0}}\sim 3$ TeV. We checked that we obtain good agreement with~\cite{Low:2014cba}, which performed the same analysis at 14 and 100 $pp$ colliders for L~=~3 ab$^{-1}$. Our sensitivities in Fig.~\ref{Fig:DisTracks} are higher because of the use of more stringent cuts.

We finally remark that, at future colliders, the reach of this channel will likely benefit by extending tracks reconstruction below the current $\sim 30$ cm value. Due to our method for estimating the background, it was not possible here to address more quantitatively this expectation.

\newpage

\section{Direct and Indirect Detection}
\label{sec:DDIDpheno}

In this Section we briefly review the constraints and perspectives for Direct and Indirect searches. We do not aim, however, at a comprehensive analysis, that we leave for upcoming work. 

We remind that the constraints considered here depend on the assumption that $\chi$ makes the whole of the DM in the Universe. If that is not the case, the bounds can be relaxed. 

\begin{figure}[t]
\centering
\includegraphics[width=0.327\textwidth]{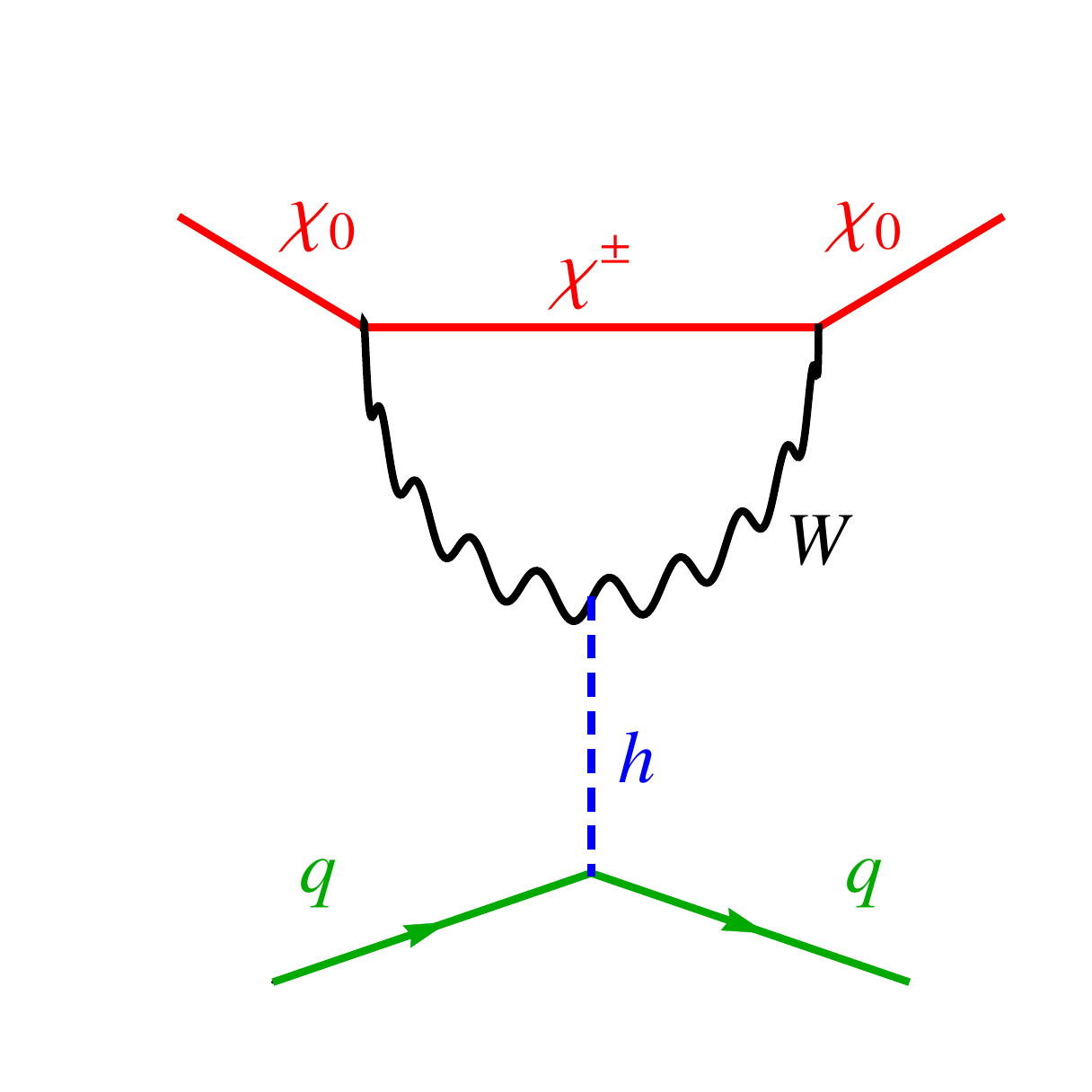} 
\includegraphics[width=0.327\textwidth]{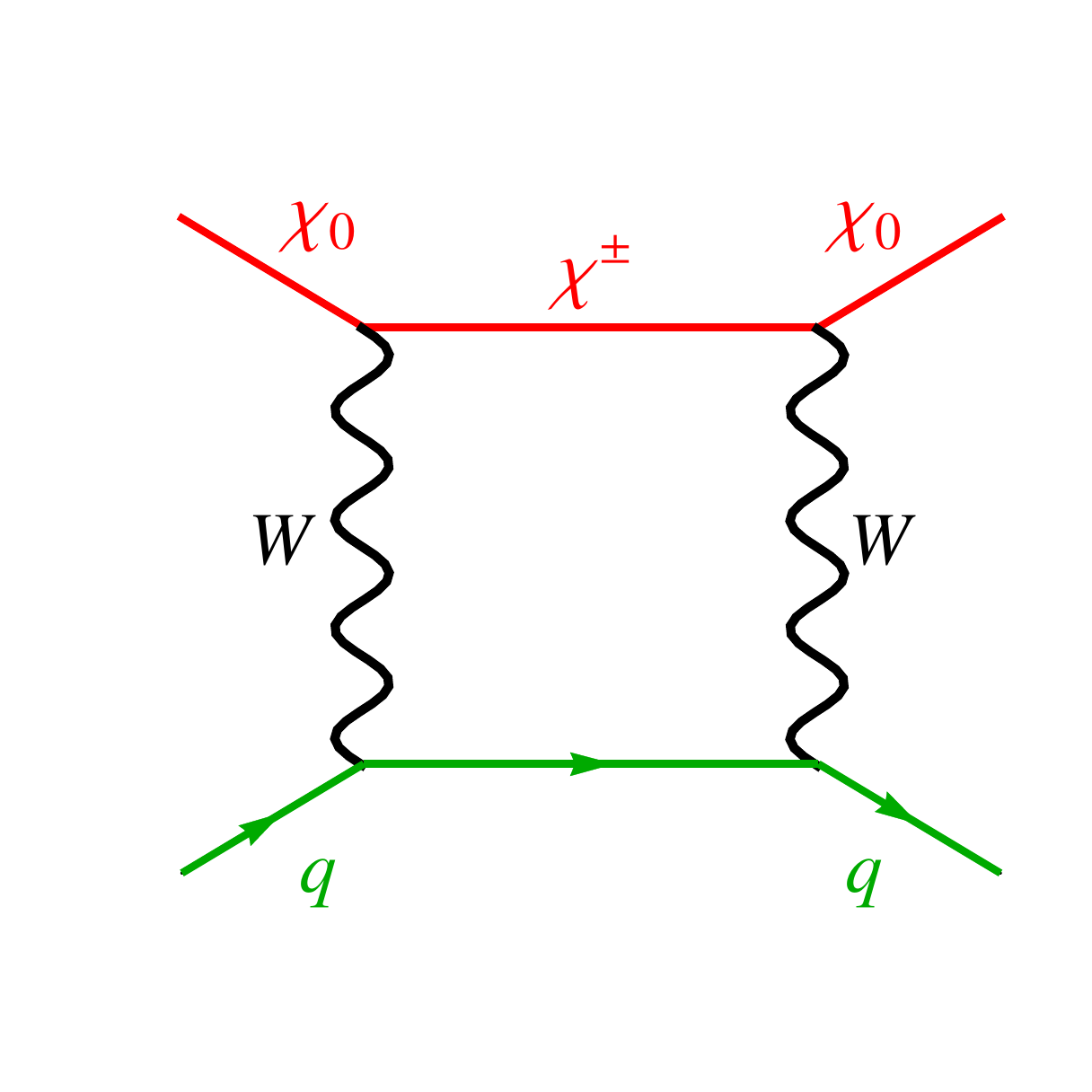} 
\includegraphics[width=0.327\textwidth]{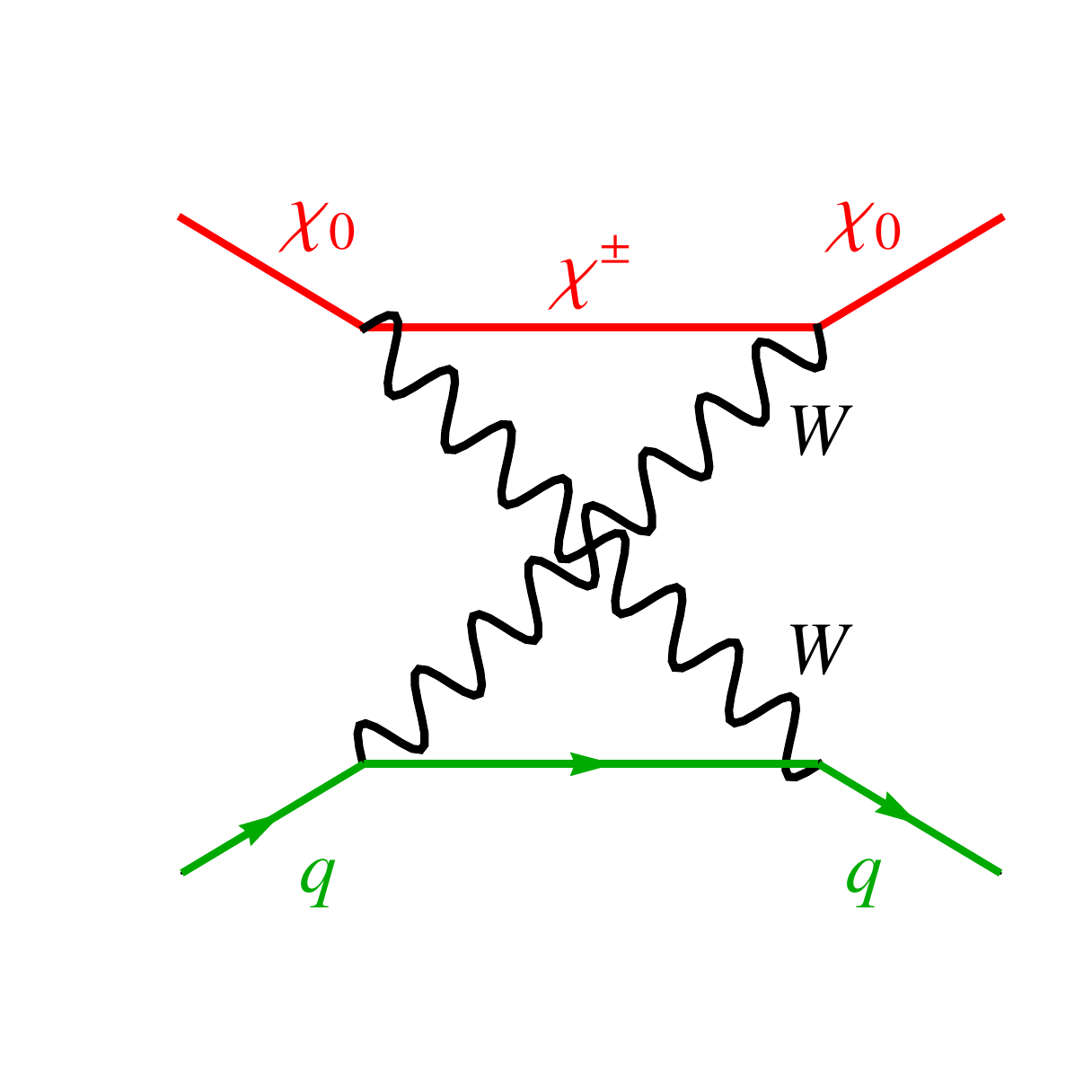} \\
\includegraphics[width=0.327\textwidth]{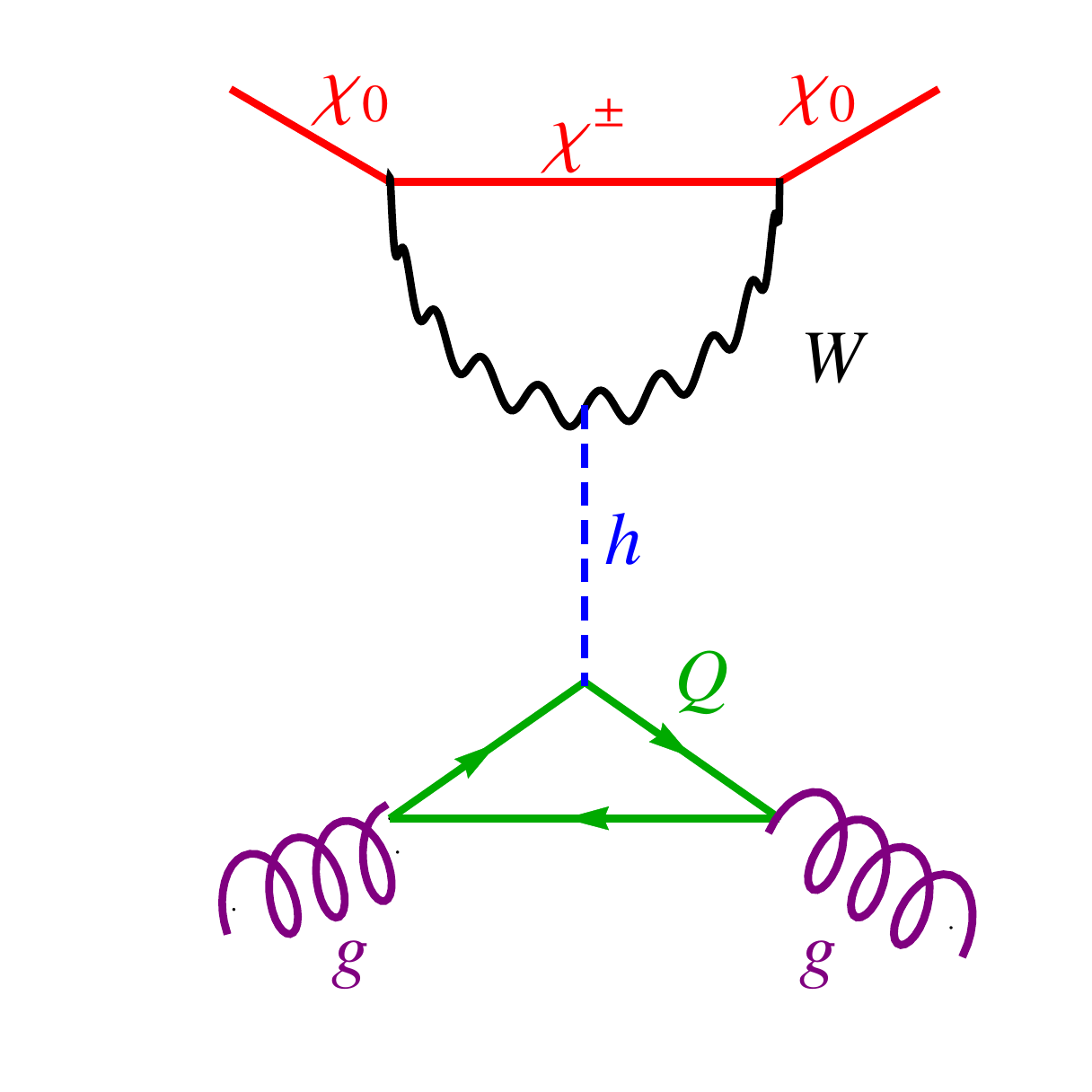} 
\includegraphics[width=0.327\textwidth]{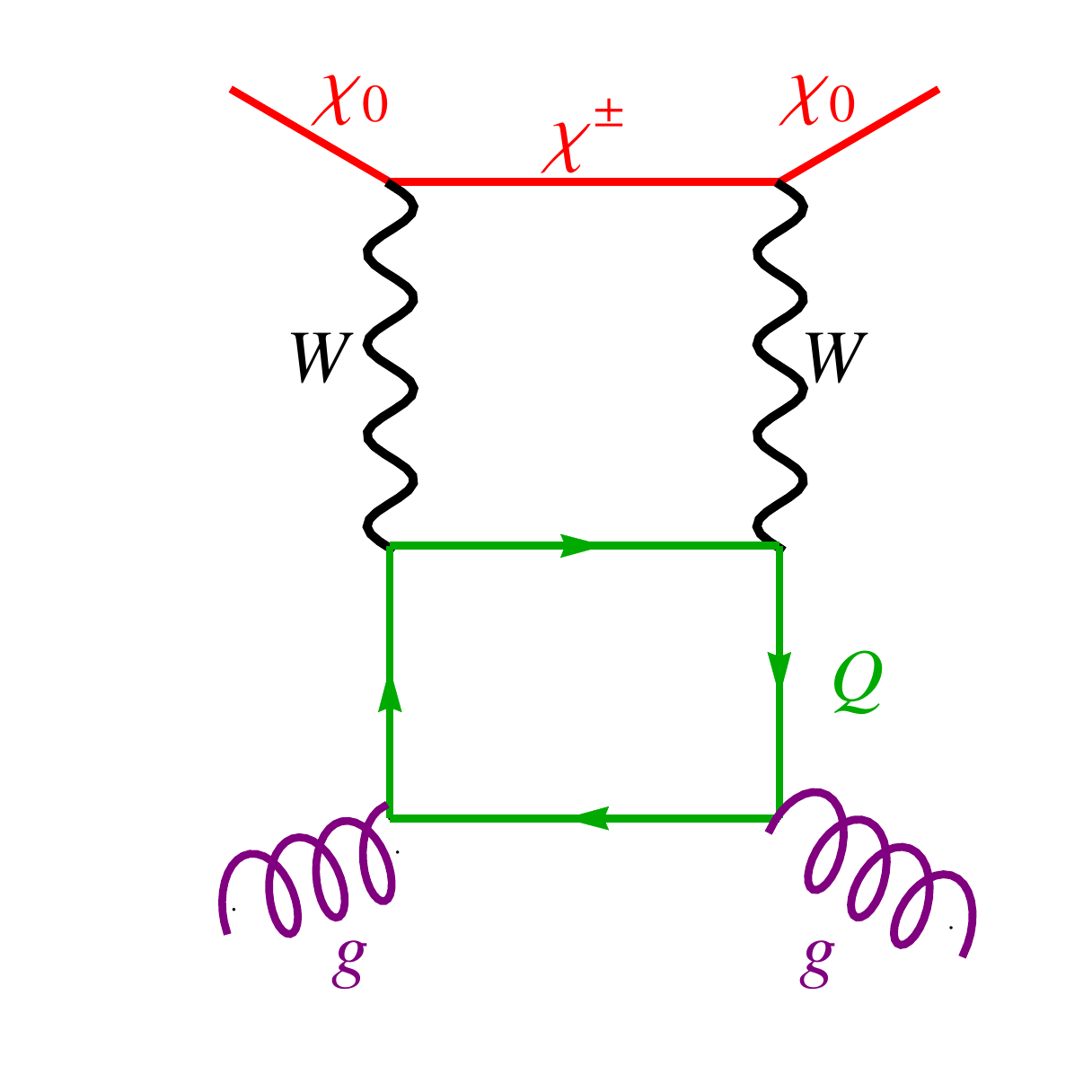} 
\includegraphics[width=0.327\textwidth]{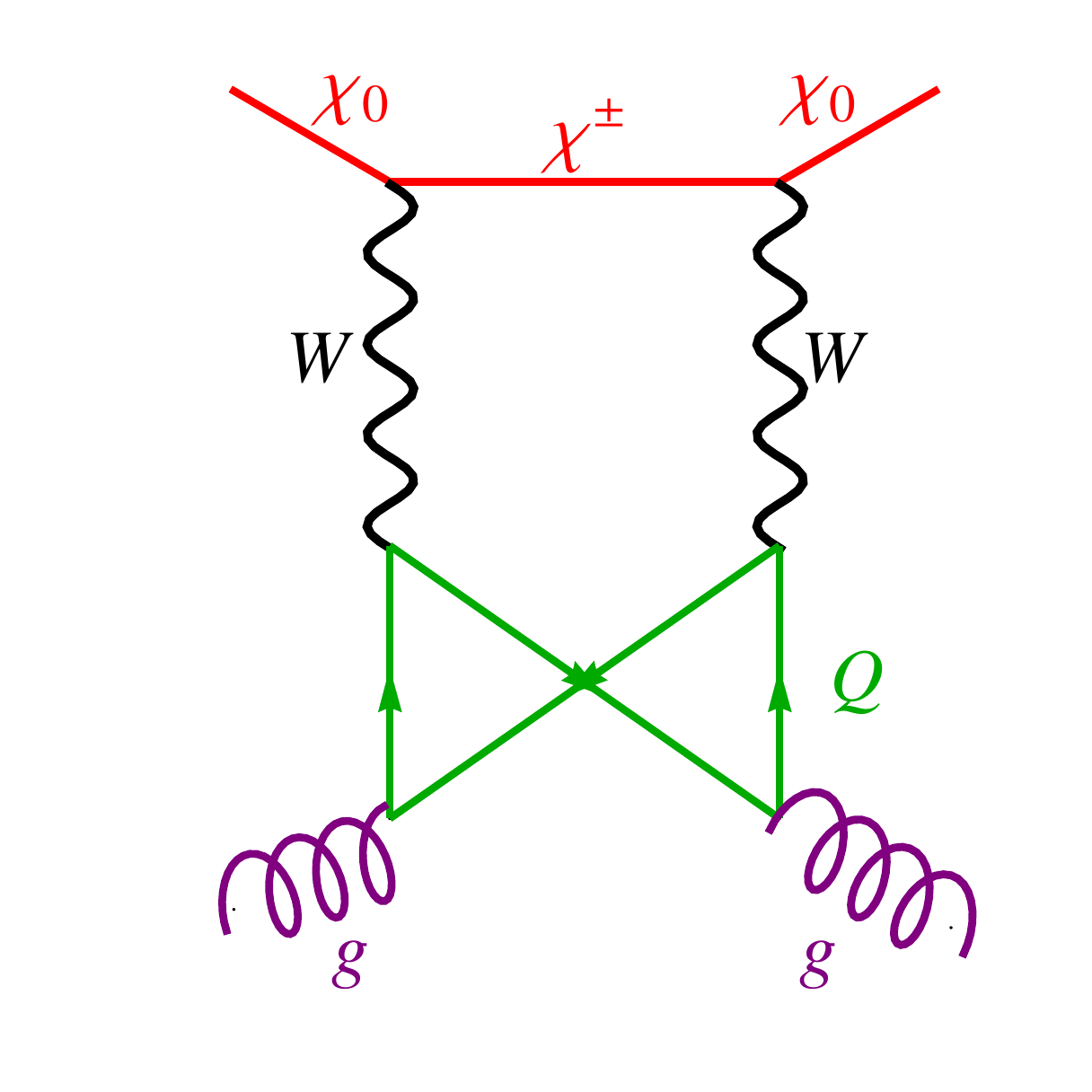} 
\caption{\label{fig:DD}Illustration of the main 1-loop and 2-loops diagrams relevant for the computation of the scattering cross section on nuclei in Direct Detection.}
\end{figure}

\paragraph{Direct Detection.} The scattering on nuclei, relevant for Direct Detection (DD), proceeds at higher loops for this candidate, since the lack of coupling with the $Z$ boson and the Higgs forbids tree level t-channel diagrams~\cite{Cirelli:2005uq}. At one loop the process proceeds via the exchange of a box of $W$ bosons or a $W$ and a Higgs. At 2-loops, the scattering with the gluons in the nucleons becomes possible via a quark loop. See fig.~\ref{fig:DD} for an illustration. 
The computation is rather involved, due to subtle cancellations which occur between different operators (notably contributing to the 1-loop diagrams, which makes the inclusion of 2-loops necessary). It has been discussed over the years in~\cite{Cirelli:2005uq,Essig:2007az,Hisano:2010ct,Hisano:2011cs,Hill:2011be,Hisano:2012wm,Farina:2013mla,DelNobile:2013sia,Hill:2013hoa,Hill:2014yka}. 
The most recent explicit computation, reported in~\cite{Hill:2013hoa} and based on~\cite{Hill:2014yka}, is performed in the framework of the `heavy WIMP effective theory' and therefore assumes $m_W,m_h \ll M_\chi$. It yields
\begin{equation}
\label{Sigma_ID}
\sigma_{\rm SI} = 1.3^{+1.3}_{-0.6} \cdot 10^{-47} {\rm cm}^2 .
\end{equation}
This value is unfortunately below the sensitivity of the current DD experiment and, for $M_\chi \gtrsim 1~{\rm TeV}$, also below the reach of the next generation~\cite{Cushman:2013zza}. For multi-TeV mass it is also dangerously close to the `WIMP discovery limit' imposed by the neutrino background. The prospects for detection via DD are therefore dim. 

\begin{figure}[t]
\centering
\includegraphics[width=0.327\textwidth]{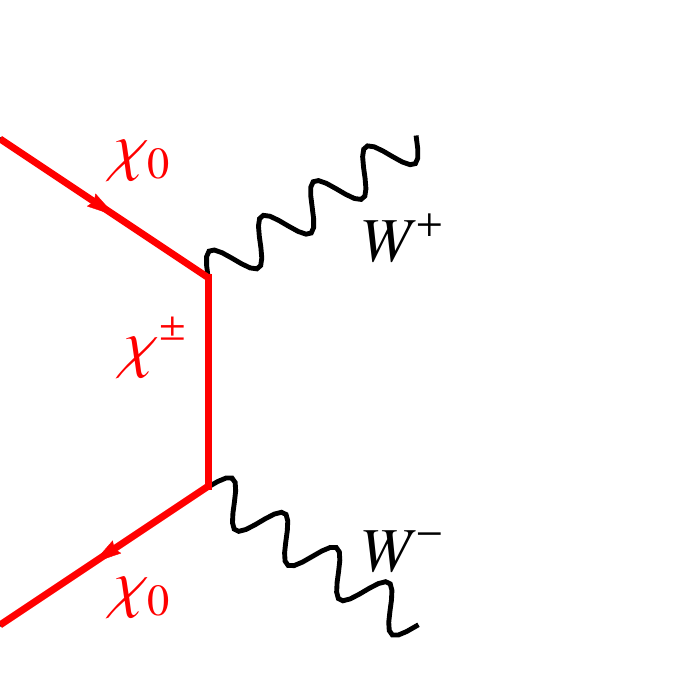} 
\includegraphics[width=0.327\textwidth]{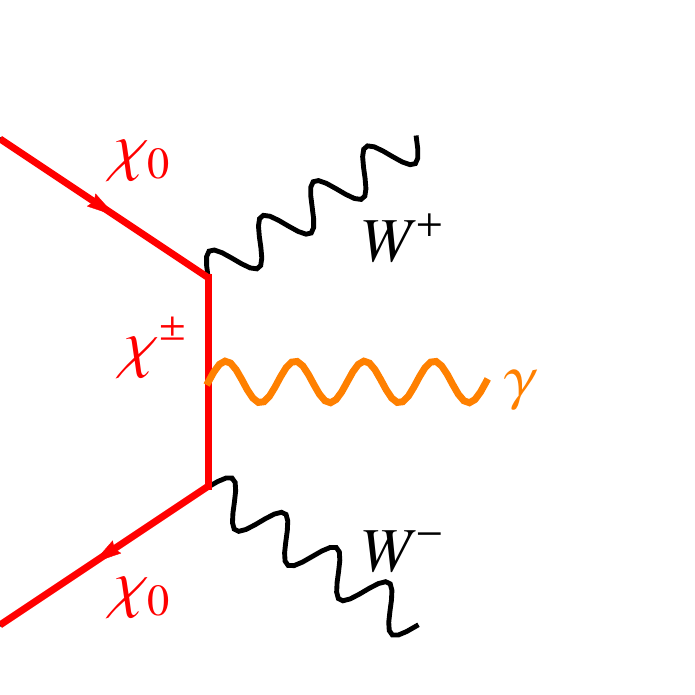} 
\includegraphics[width=0.327\textwidth]{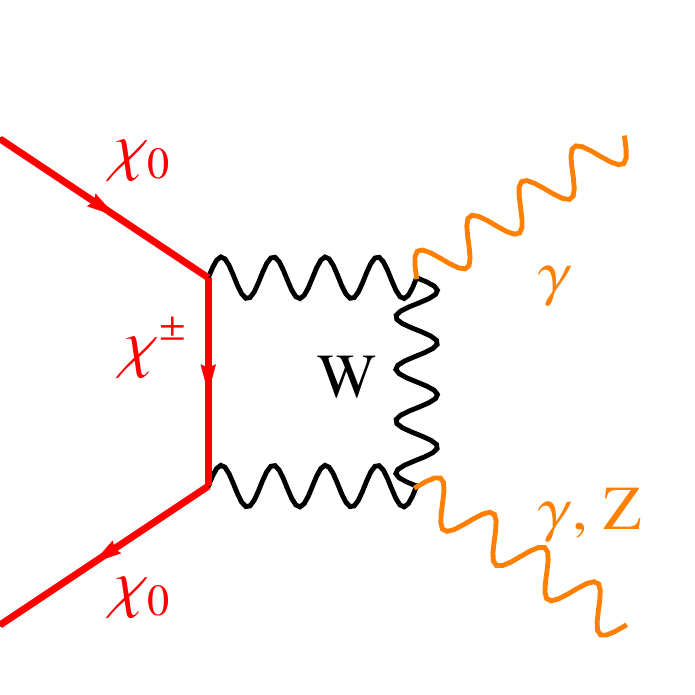} 
\caption{\label{fig:ID}Illustration of the main tree-level and 1-loop diagrams relevant for annihilation signals in Direct Detection.}
\end{figure}

\paragraph{Indirect Detection.} The DM triplet annihilates at tree level into $W^+W^-$ and into three-body states with an internally irradiated photon. At 1-loop annihilations into $\gamma\gamma$ arise (see fig.~\ref{fig:ID} for an illustration). These processes give origin to fluxes of secondary particles which would constitute exotic contributions on top of the ordinary astrophysical fluxes, and which are therefore constrained by current observations. 

Detailed analyses have been performed in~\cite{Cohen:2013ama}, \cite{Fan:2013faa} and \cite{Hryczuk:2014hpa}. According to the latter, the most relevant bounds come from antiproton and gamma-ray line measurements. Antiprotons are abundantly produced in the $WW$ DM annihilation channel and the measurements by the {\sc Pamela} satellite~\cite{Adriani:2010rc,Adriani:2012paa} significantly constrain any exotic component. However, the DM predictions are highly sensitive to the propagation model adopted for charged particles in the Galaxy, in particular to the thickness of the containment halo inside which cosmic rays diffuse. Ref.~\cite{Hryczuk:2014hpa} finds that antiprotons exclude the range $M_\chi \lesssim 1$ TeV and $1.9\ {\rm TeV} \lesssim M_\chi \lesssim 2.65$ TeV if a very thick (20 kpc) diffusive halo is assumed, while only the portions $M_\chi \lesssim 400$ GeV and 2.21 TeV $< M_\chi <$ 2.46 TeV are ruled out if the halo is as thin as 1 kpc. Both choices are probably rather unrealistic but they generously bracket the current uncertainty.

A similar situation occurs for lines (or sharp features) in the gamma-ray spectrum, originated by the rightmost two diagrams in fig.~\ref{fig:ID}. These limits are very sensitive to the choice of DM distribution profile in the Galactic Center (GC) region. Ref.~\cite{Hryczuk:2014hpa} finds that the exclusion contours from the {\sc Hess} search for lines in the GC~\cite{Abramowski:2013ax} rule out the whole range $M_\chi \lesssim 500$ GeV and $1.7\ {\rm TeV} \lesssim M_\chi \lesssim 3.5$ TeV if a benchmark Einasto profile is chosen\footnote{Both antiproton- and gamma lines- searches probe the relatively low mass region and the range around a resonance in the annihilation cross section, which is due to the Sommerfeld enhancement.}. However, if a (rather implausible) Burkert profile with a very large core is adopted, only the portion $2.25\ {\rm TeV} \lesssim M_\chi \lesssim 2.45$ TeV can be excluded. 

\medskip

In the near future, {\sc Cta} should be able to significantly improve on line searches for the GC region~\cite{Doro:2012xx,Wood:2013taa}, by probing annihilation cross sections smaller than $10^{-26} {\rm cm}^3/{\rm s}$ on most of the mass range.\footnote{However, it has been claimed that  a proper accounting of systematic uncertainties and of diffuse emission might make the task more difficult than foreseen~\cite{Silverwood:2014yza}.}
Concerning antiprotons, some improvement should come from upcoming {\sc Ams-02} data~\cite{Cirelli:2013hv}. Increased sensitivity could also come from {\sc Fermi-LAT} and {\sc Gamma-400} observation of dwarf galaxies~\cite{Bhattacherjee:2014dya}.

\section{Conclusions and outlook}
\label{sec:conclusions}

\begin{figure}[t]
\centering
\includegraphics[width=\textwidth]{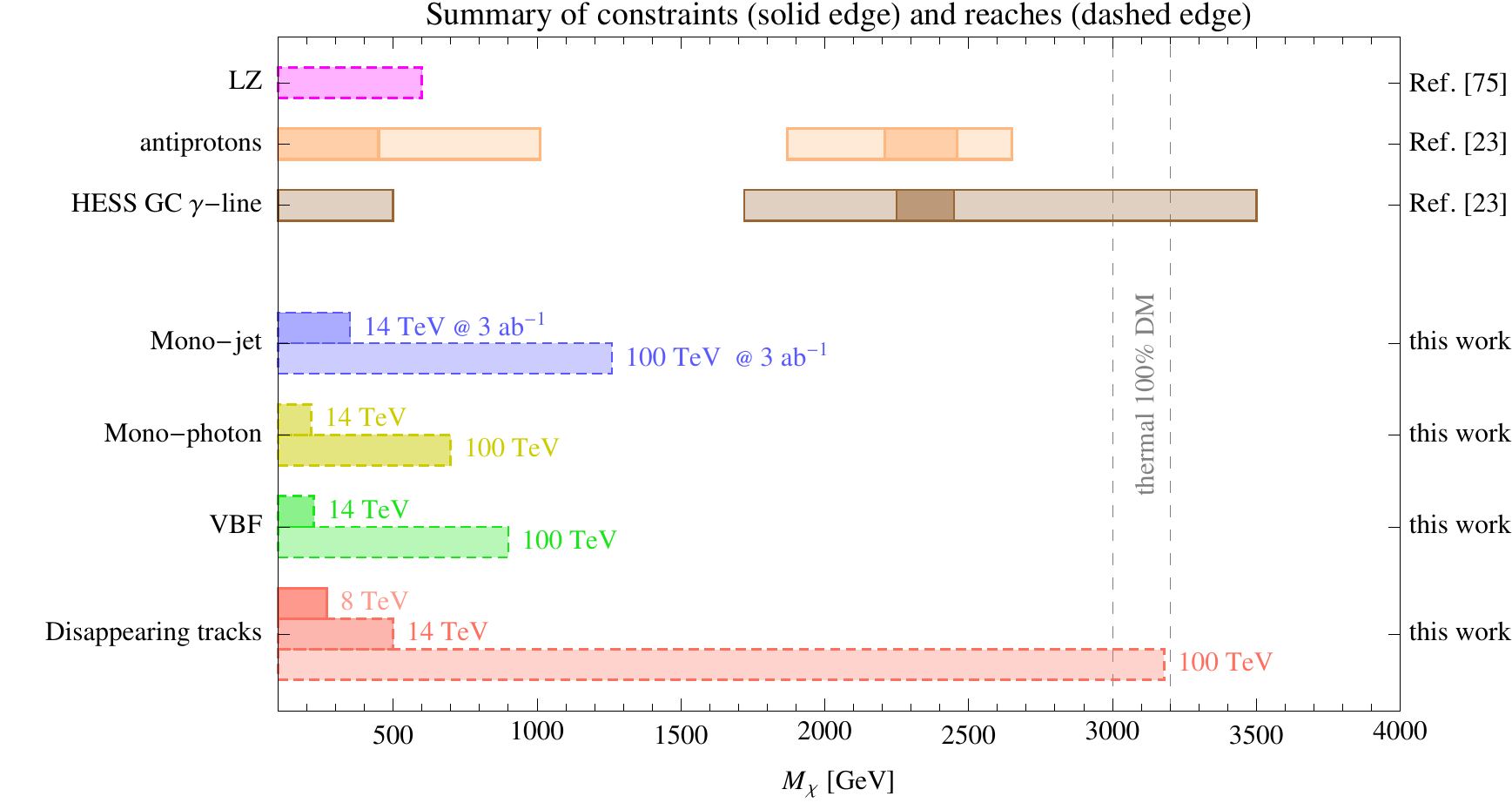} 
\caption{\label{fig:summary} Indication of the current bounds and future prospects for the electroweak triplet Dark Matter candidate. Solid contours show the current bounds. Dashed contours refer to the reach of future experiments. For the collider analysis we have considered the 95 \% CL sensitivity. For definiteness, at a 100 TeV collider we show the reach for L~=~3~ab$^{-1}$ and 1\% of background systematics. As discussed in the text, for disappearing tracks the estimate of the background at future colliders is particularly challenging. In this case, the reach refers to a moderate choice of the background uncertainty (the dashed line in Fig.~\ref{Fig:DisTracks}).}
\end{figure}

Searches of New Physics at {\sc Lhc} have been unfruitful so far. The lack of evidence of new particles and interactions at the TeV scale puts tension to {\textit{natural}} extensions of the SM, i.e. scenarios conceived to address the hierarchy problem of the electroweak scale. It is certainly premature to abandon naturalness as a criterion to approach NP. Still, it is worth considering different attitudes, for instance focussing on other open questions and investigate possible solutions in the context of NP models. 

\smallskip

In particular understanding the nature of Dark Matter is one of the most pressing challenges of modern astroparticle physics.
Here we investigate a simple solution to this problem, inspired by the Minimal Dark Matter approach \cite{Cirelli:2005uq,Cirelli:2009uv}. We consider an electroweak fermion triplet as a Dark Matter candidate. Its stability is automatic if the accidental $B-L$ symmetry of the SM, or a discrete subgroup of it, is respected by NP interactions. This particle is a prototype of a WIMP candidate and it achieves the correct relic abundance for $M_{\chi}\sim 3.0-3.2$ TeV. Different masses are also viable in presence of non-thermal production mechanisms, in non-standard cosmological scenarios or simply 
if the candidate accounts only for a fraction of the Dark Matter abundance.

\smallskip

As exposed in Section \ref{sec:motivations}, this minimal extention of the SM has additional attractive features. An electroweak triplet at the TeV scale can influence the running of the quartic coupling of the Higgs, stabilizing the Higgs vacuum. Moreover, it does not introduce large radiative corrections to the Higgs mass, and it helps to achieve the unification of the gauge couplings. This particle emerges also in more general scenarios, like SUSY models~\cite{Wells:2003tf,ArkaniHamed:2004fb,Giudice:2004tc,Arvanitaki:2012ps,Hall:2011jd,Hall:2012zp,Hall:2013eko,Hall:2014vga}, GUT constructions~\cite{Frigerio:2009wf}, and also in other contexts~\cite{Ma:2008cu,Hirsch:2013ola}.

\smallskip

Searches of this Dark Matter candidate with Direct Detection experiments are challenging, since the loop-induced scattering cross-section off nuclei is very small, well below the sensitivity of current experiments.
Indirect Detection strategies are more promising. Gamma-rays and anti-protons observations exclude the range $M_{\chi} \lesssim 1$ TeV and $1.7\ {\rm TeV} \lesssim M_{\chi} \lesssim 3.5$ TeV, although we remind that these limits are subject to large astrophysical uncertainties. Moreover they hold under the assumption that the electroweak fermion triplet accounts for all of the observed Dark Matter abundance.
Likely, new astrophysical observations will improve current Indirect Detection bounds in the near future.

\smallskip

In this work we have studied the reach of future proton colliders for the electroweak fermion triplet. We have focussed on two scenarios: {\sc Lhc} at $\sqrt{s}=$14 TeV with L~=~3~ab$^{-1}$ and a futuristic $\sqrt{s}=$ 100 TeV collider. For the latter case we have considered two benchmark luminosities, L~=~3~ab$^{-1}$ and L~=~30~ab$^{-1}$.

We have studied four channels: monojet, monophoton, VBF processes and disappearing tracks.
Disappearing tracks are the most promising probe of this scenario. At the HL {\sc Lhc-14} they will be able to test masses $M_{\chi}\lesssim 500$ GeV.
In agreement with~\cite{Low:2014cba}, we have found that a 100 TeV collider can potentially cover a range of mass up to the thermal Dark Matter one.
Among the other searches, monojet are the most powerful. The potential reach (we refer to 95 \% CL) at a 100 TeV collider is around $M_{\chi} \sim 1.3$ TeV with L~=~3~ab$^{-1}$ and $M_{\chi} \sim 1.7$ TeV with L~=~30~ab$^{-1}.$ This estimate is based on an optimistic assumption on the systematic uncertainties on the background, i.e. 1\%. We have found that for a more conservative choice, i.e. 5\%, the reach reduces significantly (around $M_{\chi} \sim 500$ GeV). Similar conclusions have been obtained for VBF and monophoton searches, with however slightly smaller reaches.

Other potential channels that could be interesting include mono-$Z$ and mono-$W$. They will be a valuable avenue for future searches, even if, for this scenario, they are not expected to have a reach better than the one of monojet \cite{Anandakrishnan:2014exa}.

\bigskip
 
We conclude summarizing our results in Fig.~\ref{fig:summary}. In this plot we compare the sensitivities of direct, indirect and collider searches. The reach of direct searches is quite modest (the future  LZ experiment could possibly cover the region $M_{\chi}\lesssim600$ GeV~\cite{Cushman:2013zza}). Indirect searches constrain either the low mass or the high mass region (the latter thanks to the presence of the Sommerfeld enhancement). 
We have found that collider searches have the potential to fill the gap, especially with disappearing tracks. Monojet, VBF and monophoton searches will provide complementary information.

\footnotesize{
\paragraph{Acknowledgements}
We thank Chiara Arina, Brando Bellazzini, Marc Besan\c con, Claude Guyot and Valerio Rossetti for useful discussions. We acknowledge the hospitality of the Institut d'Astrophysique de Paris, where part of this work was done.
Funding and research infrastructure acknowledgements: 
\begin{itemize}
\item[$\ast$] European Research Council ({\sc Erc}) under the EU Seventh Framework Programme (FP7 2007-2013)/{\sc Erc} Starting Grant (agreement n.\ 278234 --- `{\sc NewDark}' project) [work of MC, GG and MT],
\item[$\ast$] French national research agency {\sc Anr} under contract {\sc Anr} 2010 {\sc Blanc} 041301.
\end{itemize}
}
\bibliographystyle{My}
\small
\bibliography{MDMtriplet_LHC}

\providecommand{\href}[2]{#2}\begingroup\raggedright\begin{thebibliography}{10}

\bibitem{'tHooft:1979bh}
G.~'t~Hooft
{\em NATO Adv.Study Inst.Ser.B Phys.} {\bf 59} (1980)  135.

\bibitem{Yoon:2002nt}
T.~Yoon and R.~Foot, {\em Acta Phys.Polon.} {\bf B34} (2003)  2815--2842,
\href{http://arxiv.org/abs/hep-ph/0208018}{{\tt arXiv:hep-ph/0208018
  [hep-ph]}}.

\bibitem{Shaposhnikov:2007nj}
M.~Shaposhnikov
\href{http://arxiv.org/abs/0708.3550}{{\tt arXiv:0708.3550 [hep-th]}}.

\bibitem{Dubovsky:2013ira}
S.~Dubovsky, V.~Gorbenko, and M.~Mirbabayi,
\href{http://arxiv.org/abs/1305.6939}{{\tt arXiv:1305.6939 [hep-th]}}.

\bibitem{Giudice:2013yca}
G.~F. Giudice
\href{http://arxiv.org/abs/1307.7879}{{\tt arXiv:1307.7879 [hep-ph]}}.

\bibitem{Farina:2013mla}
M.~Farina, D.~Pappadopulo, and A.~Strumia,
  \href{http://dx.doi.org/10.1007/JHEP08(2013)022}{{\em JHEP} {\bf 1308} (2013)
   022},
\href{http://arxiv.org/abs/1303.7244v3}{{\tt arXiv:1303.7244v3 [hep-ph]}}.

\bibitem{Wilczek:2013lra}
F.~Wilczek \href{http://dx.doi.org/10.1088/0264-9381/30/19/193001}{{\em
  Class.Quant.Grav.} {\bf 30} (2013)  193001},
\href{http://arxiv.org/abs/1307.7376}{{\tt arXiv:1307.7376 [hep-ph]}}.

\bibitem{Busoni:2013lha}
G.~Busoni, A.~De~Simone, E.~Morgante, and A.~Riotto,
  \href{http://dx.doi.org/10.1016/j.physletb.2013.11.069}{{\em Phys.Lett.} {\bf
  B728} (2014)  412--421},
\href{http://arxiv.org/abs/1307.2253}{{\tt arXiv:1307.2253 [hep-ph]}}.

\bibitem{Busoni:2014sya}
G.~Busoni, A.~De~Simone, J.~Gramling, E.~Morgante, and A.~Riotto,
  \href{http://dx.doi.org/10.1088/1475-7516/2014/06/060}{{\em JCAP} {\bf 1406}
  (2014)  060},
\href{http://arxiv.org/abs/1402.1275}{{\tt arXiv:1402.1275 [hep-ph]}}.

\bibitem{Busoni:2014haa}
G.~Busoni, A.~De~Simone, T.~Jacques, E.~Morgante, and A.~Riotto,
\href{http://arxiv.org/abs/1405.3101}{{\tt arXiv:1405.3101 [hep-ph]}}.

\bibitem{Buchmueller:2013dya}
O.~Buchmueller, M.~J. Dolan, and C.~McCabe,
  \href{http://dx.doi.org/10.1007/JHEP01(2014)025}{{\em JHEP} {\bf 1401} (2014)
   025},
\href{http://arxiv.org/abs/1308.6799}{{\tt arXiv:1308.6799 [hep-ph]}}.

\bibitem{Shoemaker:2011vi}
I.~M. Shoemaker and L.~Vecchi,
  \href{http://dx.doi.org/10.1103/PhysRevD.86.015023}{{\em Phys.Rev.} {\bf D86}
  (2012)  015023},
\href{http://arxiv.org/abs/1112.5457}{{\tt arXiv:1112.5457 [hep-ph]}}.

\bibitem{Dreiner:2013vla}
H.~Dreiner, D.~Schmeier, and J.~Tattersall,
  \href{http://dx.doi.org/10.1209/0295-5075/102/51001}{{\em Europhys.Lett.}
  {\bf 102} (2013)  51001},
\href{http://arxiv.org/abs/1303.3348}{{\tt arXiv:1303.3348 [hep-ph]}}.

\bibitem{Chang:2013oia}
S.~Chang, R.~Edezhath, J.~Hutchinson, and M.~Luty,
  \href{http://dx.doi.org/10.1103/PhysRevD.89.015011}{{\em Phys.Rev.} {\bf D89}
  (2014)  015011},
\href{http://arxiv.org/abs/1307.8120}{{\tt arXiv:1307.8120 [hep-ph]}}.

\bibitem{An:2013xka}
H.~An, L.-T. Wang, and H.~Zhang,
  \href{http://dx.doi.org/10.1103/PhysRevD.89.115014}{{\em Phys.Rev.} {\bf D89}
  (2014)  115014},
\href{http://arxiv.org/abs/1308.0592}{{\tt arXiv:1308.0592 [hep-ph]}}.

\bibitem{Bai:2013iqa}
Y.~Bai and J.~Berger, \href{http://dx.doi.org/10.1007/JHEP11(2013)171}{{\em
  JHEP} {\bf 1311} (2013)  171},
\href{http://arxiv.org/abs/1308.0612}{{\tt arXiv:1308.0612 [hep-ph]}}.

\bibitem{DiFranzo:2013vra}
A.~DiFranzo, K.~I. Nagao, A.~Rajaraman, and T.~M.~P. Tait,
  \href{http://dx.doi.org/10.1007/JHEP11(2013)014}{{\em JHEP} {\bf 1311} (2013)
   014},
\href{http://arxiv.org/abs/1308.2679}{{\tt arXiv:1308.2679 [hep-ph]}}.

\bibitem{Papucci:2014iwa}
M.~Papucci, A.~Vichi, and K.~M. Zurek,
\href{http://arxiv.org/abs/1402.2285}{{\tt arXiv:1402.2285 [hep-ph]}}.

\bibitem{deSimone:2014pda}
A.~De~Simone, G.~F. Giudice, and A.~Strumia,
  \href{http://dx.doi.org/10.1007/JHEP06(2014)081}{{\em JHEP} {\bf 1406} (2014)
   081},
\href{http://arxiv.org/abs/1402.6287}{{\tt arXiv:1402.6287 [hep-ph]}}.

\bibitem{Ibe:2012sx}
M.~Ibe, S.~Matsumoto, and R.~Sato,
  \href{http://dx.doi.org/10.1016/j.physletb.2013.03.015}{{\em Phys.Lett.} {\bf
  B721} (2013)  252--260},
\href{http://arxiv.org/abs/1212.5989}{{\tt arXiv:1212.5989 [hep-ph]}}.

\bibitem{Cirelli:2005uq}
M.~Cirelli, N.~Fornengo, and A.~Strumia,
  \href{http://dx.doi.org/10.1016/j.nuclphysb.2006.07.012}{{\em Nucl.Phys.}
  {\bf B753} (2006)  178--194},
\href{http://arxiv.org/abs/hep-ph/0512090}{{\tt arXiv:hep-ph/0512090
  [hep-ph]}}.

\bibitem{Cirelli:2009uv}
M.~Cirelli and A.~Strumia,
  \href{http://dx.doi.org/10.1088/1367-2630/11/10/105005}{{\em New J.Phys.}
  {\bf 11} (2009)  105005},
\href{http://arxiv.org/abs/0903.3381}{{\tt arXiv:0903.3381 [hep-ph]}}.

\bibitem{Hryczuk:2014hpa}
A.~Hryczuk, I.~Cholis, R.~Iengo, M.~Tavakoli, and P.~Ullio,
\href{http://arxiv.org/abs/1401.6212}{{\tt arXiv:1401.6212 [astro-ph.HE]}}.

\bibitem{Chao:2012mx}
W.~Chao, M.~Gonderinger, and M.~J. Ramsey-Musolf,
  \href{http://dx.doi.org/10.1103/PhysRevD.86.113017}{{\em Phys.Rev.} {\bf D86}
  (2012)  113017},
\href{http://arxiv.org/abs/1210.0491}{{\tt arXiv:1210.0491 [hep-ph]}}.

\bibitem{Ade:2014xna}
{\bf BICEP2 Collaboration}, P.~Ade {\em et al.,}
  \href{http://dx.doi.org/10.1103/PhysRevLett.112.241101}{{\em Phys.Rev.Lett.}
  {\bf 112} (2014)  241101},
\href{http://arxiv.org/abs/1403.3985}{{\tt arXiv:1403.3985 [astro-ph.CO]}}.

\bibitem{Espinosa:2007qp}
J.~Espinosa, G.~Giudice, and A.~Riotto,
  \href{http://dx.doi.org/10.1088/1475-7516/2008/05/002}{{\em JCAP} {\bf 0805}
  (2008)  002},
\href{http://arxiv.org/abs/0710.2484}{{\tt arXiv:0710.2484 [hep-ph]}}.

\bibitem{Kobakhidze:2013tn}
A.~Kobakhidze and A.~Spencer-Smith,
  \href{http://dx.doi.org/10.1016/j.physletb.2013.04.013}{{\em Phys.Lett.} {\bf
  B722} (2013)  130--134},
\href{http://arxiv.org/abs/1301.2846}{{\tt arXiv:1301.2846 [hep-ph]}}.

\bibitem{Fairbairn:2014zia}
M.~Fairbairn and R.~Hogan,
\href{http://arxiv.org/abs/1403.6786}{{\tt arXiv:1403.6786 [hep-ph]}}.

\bibitem{Enqvist:2014bua}
K.~Enqvist, T.~Meriniemi, and S.~Nurmi,
\href{http://arxiv.org/abs/1404.3699}{{\tt arXiv:1404.3699 [hep-ph]}}.

\bibitem{Kobakhidze:2014xda}
A.~Kobakhidze and A.~Spencer-Smith,
\href{http://arxiv.org/abs/1404.4709}{{\tt arXiv:1404.4709 [hep-ph]}}.

\bibitem{Hook:2014uia}
A.~Hook, J.~Kearney, B.~Shakya, and K.~M. Zurek,
\href{http://arxiv.org/abs/1404.5953}{{\tt arXiv:1404.5953 [hep-ph]}}.

\bibitem{EliasMiro:2012ay}
J.~Elias-Miro, J.~R. Espinosa, G.~F. Giudice, H.~M. Lee, and A.~Strumia,
  \href{http://dx.doi.org/10.1007/JHEP06(2012)031}{{\em JHEP} {\bf 1206} (2012)
   031},
\href{http://arxiv.org/abs/1203.0237}{{\tt arXiv:1203.0237 [hep-ph]}}.

\bibitem{Giudice:2004tc}
G.~Giudice and A.~Romanino,
  \href{http://dx.doi.org/10.1016/j.nuclphysb.2004.11.048}{{\em Nucl.Phys.}
  {\bf B699} (2004)  65--89},
\href{http://arxiv.org/abs/hep-ph/0406088}{{\tt arXiv:hep-ph/0406088
  [hep-ph]}}.

\bibitem{Frigerio:2009wf}
M.~Frigerio and T.~Hambye,
  \href{http://dx.doi.org/10.1103/PhysRevD.81.075002}{{\em Phys.Rev.} {\bf D81}
  (2010)  075002},
\href{http://arxiv.org/abs/0912.1545}{{\tt arXiv:0912.1545 [hep-ph]}}.

\bibitem{Mahbubani:2005pt}
R.~Mahbubani and L.~Senatore,
  \href{http://dx.doi.org/10.1103/PhysRevD.73.043510}{{\em Phys.Rev.} {\bf D73}
  (2006)  043510},
\href{http://arxiv.org/abs/hep-ph/0510064}{{\tt arXiv:hep-ph/0510064
  [hep-ph]}}.

\bibitem{Wells:2003tf}
J.~D. Wells
\href{http://arxiv.org/abs/hep-ph/0306127}{{\tt arXiv:hep-ph/0306127
  [hep-ph]}}.

\bibitem{ArkaniHamed:2004fb}
N.~Arkani-Hamed and S.~Dimopoulos,
  \href{http://dx.doi.org/10.1088/1126-6708/2005/06/073}{{\em JHEP} {\bf 0506}
  (2005)  073},
\href{http://arxiv.org/abs/hep-th/0405159}{{\tt arXiv:hep-th/0405159
  [hep-th]}}.

\bibitem{Arvanitaki:2012ps}
A.~Arvanitaki, N.~Craig, S.~Dimopoulos, and G.~Villadoro,
  \href{http://dx.doi.org/10.1007/JHEP02(2013)126}{{\em JHEP} {\bf 1302} (2013)
   126},
\href{http://arxiv.org/abs/1210.0555}{{\tt arXiv:1210.0555 [hep-ph]}}.

\bibitem{Hall:2011jd}
L.~J. Hall and Y.~Nomura, \href{http://dx.doi.org/10.1007/JHEP01(2012)082}{{\em
  JHEP} {\bf 1201} (2012)  082},
\href{http://arxiv.org/abs/1111.4519}{{\tt arXiv:1111.4519 [hep-ph]}}.

\bibitem{Hall:2012zp}
L.~J. Hall, Y.~Nomura, and S.~Shirai,
  \href{http://dx.doi.org/10.1007/JHEP01(2013)036}{{\em JHEP} {\bf 1301} (2013)
   036},
\href{http://arxiv.org/abs/1210.2395}{{\tt arXiv:1210.2395 [hep-ph]}}.

\bibitem{Hall:2013eko}
L.~J. Hall and Y.~Nomura, \href{http://dx.doi.org/10.1007/JHEP02(2014)129}{{\em
  JHEP} {\bf 1402} (2014)  129},
\href{http://arxiv.org/abs/1312.6695}{{\tt arXiv:1312.6695 [hep-ph]}}.

\bibitem{Hall:2014vga}
L.~J. Hall, Y.~Nomura, and S.~Shirai,
\href{http://arxiv.org/abs/1403.8138}{{\tt arXiv:1403.8138 [hep-ph]}}.

\bibitem{Aad:2013yna}
{\bf ATLAS Collaboration}, G.~Aad {\em et al.,}
  \href{http://dx.doi.org/10.1103/PhysRevD.88.112006}{{\em Phys.Rev.} {\bf D88}
  (2013)  112006},
\href{http://arxiv.org/abs/1310.3675}{{\tt arXiv:1310.3675 [hep-ex]}}.

\bibitem{Cohen:2014hxa}
T.~Cohen, R.~T. D'Agnolo, M.~Hance, H.~K. Lou, and J.~G. Wacker,
\href{http://arxiv.org/abs/1406.4512}{{\tt arXiv:1406.4512 [hep-ph]}}.

\bibitem{Alloul:2013bka}
A.~Alloul, N.~D. Christensen, C.~Degrande, C.~Duhr, and B.~Fuks,
  \href{http://dx.doi.org/10.1016/j.cpc.2014.04.012}{{\em Comput.Phys.Commun.}
  {\bf 185} (2014)  2250--2300},
\href{http://arxiv.org/abs/1310.1921}{{\tt arXiv:1310.1921 [hep-ph]}}.

\bibitem{Alwall:2011uj}
J.~Alwall, M.~Herquet, F.~Maltoni, O.~Mattelaer, and T.~Stelzer,
  \href{http://dx.doi.org/10.1007/JHEP06(2011)128}{{\em JHEP} {\bf 1106} (2011)
   128},
\href{http://arxiv.org/abs/1106.0522}{{\tt arXiv:1106.0522 [hep-ph]}}.

\bibitem{Sjostrand:2006za}
T.~Sjostrand, S.~Mrenna, and P.~Z. Skands,
  \href{http://dx.doi.org/10.1088/1126-6708/2006/05/026}{{\em JHEP} {\bf 0605}
  (2006)  026},
\href{http://arxiv.org/abs/hep-ph/0603175}{{\tt arXiv:hep-ph/0603175
  [hep-ph]}}.

\bibitem{Hook:2014rka}
A.~Hook and A.~Katz,
\href{http://arxiv.org/abs/1407.2607}{{\tt arXiv:1407.2607 [hep-ph]}}.

\bibitem{deFavereau:2013fsa}
{\bf DELPHES 3}, J.~de~Favereau {\em et al.,}
  \href{http://dx.doi.org/10.1007/JHEP02(2014)057}{{\em JHEP} {\bf 1402} (2014)
   057},
\href{http://arxiv.org/abs/1307.6346}{{\tt arXiv:1307.6346 [hep-ex]}}.

\bibitem{pgs}
 \url{http://www.physics.ucdavis.edu/~conway/research/software/pgs/pgs4-general.htm}.

\bibitem{CMS-PAS-EXO-12-048}
{\bf CMS Collaboration} Tech. Rep. CMS-PAS-EXO-12-048, CERN, Geneva, 2013.

\bibitem{ATLAS-CONF-2012-147}
{\bf ATLAS Collaboration} Tech. Rep. ATLAS-CONF-2012-147, CERN, Geneva, Nov,
  2012.

\bibitem{CMS-PAS-EXO-12-047}
{\bf CMS Collaboration} Tech. Rep. CMS-PAS-EXO-12-047, CERN, Geneva, 2014.

\bibitem{Aad:2012fw}
{\bf ATLAS Collaboration}, G.~Aad {\em et al.,}
  \href{http://dx.doi.org/10.1103/PhysRevLett.110.011802}{{\em Phys.Rev.Lett.}
  {\bf 110} (2013)  011802},
\href{http://arxiv.org/abs/1209.4625}{{\tt arXiv:1209.4625 [hep-ex]}}.

\bibitem{Chatrchyan:2014tja}
{\bf CMS Collaboration}, S.~Chatrchyan {\em et al.,}
\href{http://arxiv.org/abs/1404.1344}{{\tt arXiv:1404.1344 [hep-ex]}}.

\bibitem{Low:2014cba}
M.~Low and L.-T. Wang,
\href{http://arxiv.org/abs/1404.0682}{{\tt arXiv:1404.0682 [hep-ph]}}.

\bibitem{Fox:2011pm}
P.~J. Fox, R.~Harnik, J.~Kopp, and Y.~Tsai,
  \href{http://dx.doi.org/10.1103/PhysRevD.85.056011}{{\em Phys.Rev.} {\bf D85}
  (2012)  056011},
\href{http://arxiv.org/abs/1109.4398}{{\tt arXiv:1109.4398 [hep-ph]}}.

\bibitem{Hagiwara:2012we}
K.~Hagiwara, D.~Marfatia, and T.~Yamada,
  \href{http://dx.doi.org/10.1103/PhysRevD.89.094017}{{\em Phys.Rev.} {\bf D89}
  (2014)  094017},
\href{http://arxiv.org/abs/1207.6857}{{\tt arXiv:1207.6857 [hep-ph]}}.

\bibitem{Delannoy:2013ata}
A.~G. Delannoy, B.~Dutta, A.~Gurrola, W.~Johns, T.~Kamon, {\em et al.,}
  \href{http://dx.doi.org/10.1103/PhysRevLett.111.061801}{{\em Phys.Rev.Lett.}
  {\bf 111} (2013)  061801},
\href{http://arxiv.org/abs/1304.7779}{{\tt arXiv:1304.7779 [hep-ph]}}.

\bibitem{Feng:1994mq}
J.~L. Feng and M.~J. Strassler,
  \href{http://dx.doi.org/10.1103/PhysRevD.51.4661}{{\em Phys.Rev.} {\bf D51}
  (1995)  4661--4694},
\href{http://arxiv.org/abs/hep-ph/9408359}{{\tt arXiv:hep-ph/9408359
  [hep-ph]}}.

\bibitem{Feng:1999fu}
J.~L. Feng, T.~Moroi, L.~Randall, M.~Strassler, and S.-f. Su,
  \href{http://dx.doi.org/10.1103/PhysRevLett.83.1731}{{\em Phys.Rev.Lett.}
  {\bf 83} (1999)  1731--1734},
\href{http://arxiv.org/abs/hep-ph/9904250}{{\tt arXiv:hep-ph/9904250
  [hep-ph]}}.

\bibitem{Gunion:1999jr}
J.~F. Gunion and S.~Mrenna,
  \href{http://dx.doi.org/10.1103/PhysRevD.62.015002}{{\em Phys.Rev.} {\bf D62}
  (2000)  015002},
\href{http://arxiv.org/abs/hep-ph/9906270}{{\tt arXiv:hep-ph/9906270
  [hep-ph]}}.

\bibitem{Gunion:2001fu}
J.~F. Gunion and S.~Mrenna,
  \href{http://dx.doi.org/10.1103/PhysRevD.64.075002}{{\em Phys.Rev.} {\bf D64}
  (2001)  075002},
\href{http://arxiv.org/abs/hep-ph/0103167}{{\tt arXiv:hep-ph/0103167
  [hep-ph]}}.

\bibitem{Barr:2002ex}
A.~Barr, C.~Lester, M.~A. Parker, B.~Allanach, and P.~Richardson,
  \href{http://dx.doi.org/10.1088/1126-6708/2003/03/045}{{\em JHEP} {\bf 0303}
  (2003)  045},
\href{http://arxiv.org/abs/hep-ph/0208214}{{\tt arXiv:hep-ph/0208214
  [hep-ph]}}.

\bibitem{Ibe:2006de}
M.~Ibe, T.~Moroi, and T.~Yanagida,
  \href{http://dx.doi.org/10.1016/j.physletb.2006.11.061}{{\em Phys.Lett.} {\bf
  B644} (2007)  355--360},
\href{http://arxiv.org/abs/hep-ph/0610277}{{\tt arXiv:hep-ph/0610277
  [hep-ph]}}.

\bibitem{Buckley:2009kv}
M.~R. Buckley, L.~Randall, and B.~Shuve,
  \href{http://dx.doi.org/10.1007/JHEP05(2011)097}{{\em JHEP} {\bf 1105} (2011)
   097},
\href{http://arxiv.org/abs/0909.4549}{{\tt arXiv:0909.4549 [hep-ph]}}.

\bibitem{Kane:2012aa}
G.~Kane, R.~Lu, and B.~Zheng,
\href{http://arxiv.org/abs/1202.4448}{{\tt arXiv:1202.4448 [hep-ph]}}.

\bibitem{Essig:2007az}
R.~Essig \href{http://dx.doi.org/10.1103/PhysRevD.78.015004}{{\em Phys.Rev.}
  {\bf D78} (2008)  015004},
\href{http://arxiv.org/abs/0710.1668}{{\tt arXiv:0710.1668 [hep-ph]}}.

\bibitem{Hisano:2010ct}
J.~Hisano, K.~Ishiwata, and N.~Nagata,
  \href{http://dx.doi.org/10.1103/PhysRevD.82.115007}{{\em Phys.Rev.} {\bf D82}
  (2010)  115007},
\href{http://arxiv.org/abs/1007.2601}{{\tt arXiv:1007.2601 [hep-ph]}}.

\bibitem{Hisano:2011cs}
J.~Hisano, K.~Ishiwata, N.~Nagata, and T.~Takesako,
  \href{http://dx.doi.org/10.1007/JHEP07(2011)005}{{\em JHEP} {\bf 1107} (2011)
   005},
\href{http://arxiv.org/abs/1104.0228}{{\tt arXiv:1104.0228 [hep-ph]}}.

\bibitem{Hill:2011be}
R.~J. Hill and M.~P. Solon,
  \href{http://dx.doi.org/10.1016/j.physletb.2012.01.013}{{\em Phys.Lett.} {\bf
  B707} (2012)  539--545},
\href{http://arxiv.org/abs/1111.0016}{{\tt arXiv:1111.0016 [hep-ph]}}.

\bibitem{Hisano:2012wm}
J.~Hisano, K.~Ishiwata, and N.~Nagata,
  \href{http://dx.doi.org/10.1103/PhysRevD.87.035020}{{\em Phys.Rev.} {\bf D87}
  (2013)  035020},
\href{http://arxiv.org/abs/1210.5985}{{\tt arXiv:1210.5985 [hep-ph]}}.

\bibitem{DelNobile:2013sia}
M.~Cirelli, E.~Del~Nobile, and P.~Panci,
  \href{http://dx.doi.org/10.1088/1475-7516/2013/10/019}{{\em JCAP} {\bf 1310}
  (2013)  019},
\href{http://arxiv.org/abs/1307.5955}{{\tt arXiv:1307.5955 [hep-ph]}}.

\bibitem{Hill:2013hoa}
R.~J. Hill and M.~P. Solon,
  \href{http://dx.doi.org/10.1103/PhysRevLett.112.211602}{{\em Phys.Rev.Lett.}
  {\bf 112} (2014)  211602},
\href{http://arxiv.org/abs/1309.4092}{{\tt arXiv:1309.4092 [hep-ph]}}.

\bibitem{Hill:2014yka}
R.~J. Hill and M.~P. Solon,
\href{http://arxiv.org/abs/1401.3339}{{\tt arXiv:1401.3339 [hep-ph]}}.

\bibitem{Cushman:2013zza}
P.~Cushman, C.~Galbiati, D.~McKinsey, H.~Robertson, T.~Tait, {\em et al.,}
\href{http://arxiv.org/abs/1310.8327}{{\tt arXiv:1310.8327 [hep-ex]}}.

\bibitem{Cohen:2013ama}
T.~Cohen, M.~Lisanti, A.~Pierce, and T.~R. Slatyer,
  \href{http://dx.doi.org/10.1088/1475-7516/2013/10/061}{{\em JCAP} {\bf 1310}
  (2013)  061},
\href{http://arxiv.org/abs/1307.4082}{{\tt arXiv:1307.4082}}.

\bibitem{Fan:2013faa}
J.~Fan and M.~Reece, \href{http://dx.doi.org/10.1007/JHEP10(2013)124}{{\em
  JHEP} {\bf 1310} (2013)  124},
\href{http://arxiv.org/abs/1307.4400}{{\tt arXiv:1307.4400 [hep-ph]}}.

\bibitem{Adriani:2010rc}
{\bf PAMELA Collaboration}, O.~Adriani {\em et al.,}
  \href{http://dx.doi.org/10.1103/PhysRevLett.105.121101}{{\em Phys.Rev.Lett.}
  {\bf 105} (2010)  121101},
\href{http://arxiv.org/abs/1007.0821}{{\tt arXiv:1007.0821 [astro-ph.HE]}}.

\bibitem{Adriani:2012paa}
O.~Adriani, G.~Bazilevskaya, G.~Barbarino, R.~Bellotti, M.~Boezio, {\em et
  al.,}
\href{http://dx.doi.org/10.1134/S002136401222002X}{{\em JETP Lett.} {\bf 96}
  (2013)  621--627}.

\bibitem{Abramowski:2013ax}
{\bf H.E.S.S. Collaboration}, A.~Abramowski {\em et al.,}
  \href{http://dx.doi.org/10.1103/PhysRevLett.110.041301}{{\em Phys.Rev.Lett.}
  {\bf 110} (2013)  041301},
\href{http://arxiv.org/abs/1301.1173}{{\tt arXiv:1301.1173 [astro-ph.HE]}}.

\bibitem{Doro:2012xx}
{\bf CTA collaboration}, M.~Doro {\em et al.,}
  \href{http://dx.doi.org/10.1016/j.astropartphys.2012.08.002}{{\em
  Astropart.Phys.} {\bf 43} (2013)  189--214},
\href{http://arxiv.org/abs/1208.5356}{{\tt arXiv:1208.5356 [astro-ph.IM]}}.

\bibitem{Wood:2013taa}
M.~Wood, J.~Buckley, S.~Digel, S.~Funk, D.~Nieto, {\em et al.,}
\href{http://arxiv.org/abs/1305.0302}{{\tt arXiv:1305.0302 [astro-ph.HE]}}.

\bibitem{Silverwood:2014yza}
H.~Silverwood, C.~Weniger, P.~Scott, and G.~Bertone,
\href{http://arxiv.org/abs/1408.4131}{{\tt arXiv:1408.4131 [astro-ph.HE]}}.

\bibitem{Cirelli:2013hv}
M.~Cirelli and G.~Giesen,
  \href{http://dx.doi.org/10.1088/1475-7516/2013/04/015}{{\em JCAP} {\bf 1304}
  (2013)  015},
\href{http://arxiv.org/abs/1301.7079}{{\tt arXiv:1301.7079 [hep-ph]}}.

\bibitem{Bhattacherjee:2014dya}
B.~Bhattacherjee, M.~Ibe, K.~Ichikawa, S.~Matsumoto, and K.~Nishiyama,
\href{http://arxiv.org/abs/1405.4914}{{\tt arXiv:1405.4914 [hep-ph]}}.

\bibitem{Ma:2008cu}
E.~Ma and D.~Suematsu, \href{http://dx.doi.org/10.1142/S021773230903059X}{{\em
  Mod.Phys.Lett.} {\bf A24} (2009)  583--589},
\href{http://arxiv.org/abs/0809.0942}{{\tt arXiv:0809.0942 [hep-ph]}}.

\bibitem{Hirsch:2013ola}
M.~Hirsch, R.~Lineros, S.~Morisi, J.~Palacio, N.~Rojas, {\em et al.,}
  \href{http://dx.doi.org/10.1007/JHEP10(2013)149}{{\em JHEP} {\bf 1310} (2013)
   149},
\href{http://arxiv.org/abs/1307.8134}{{\tt arXiv:1307.8134 [hep-ph]}}.

\bibitem{Anandakrishnan:2014exa}
A.~Anandakrishnan, L.~M. Carpenter, and S.~Raby,
\href{http://arxiv.org/abs/1407.1833}{{\tt arXiv:1407.1833 [hep-ph]}}.

\end{thebibliography}\endgroup

%
\end{document}